\documentclass{aa}

\usepackage{graphicx}
\usepackage{txfonts}
\usepackage[colorlinks=true,allcolors=blue]{hyperref}
\usepackage{xcolor}
\usepackage{commath}
\usepackage{array}
\usepackage{subcaption}
\usepackage{soul}
\usepackage{booktabs}
\usepackage{multirow}
\def\DeltaIn{\delta}
\def\DeltaTrans{10^{-5}}
\def\Pm{\langle P_1 \rangle}
\def\Pms{\langle P_2 \rangle}
\def\lnL{\ln \mathcal{L}}
\def\Lm{\lambda_{1,0}}

\begin{document}

\title{Constraining the orbit of the retrograde planet \\ in the $\nu$ Octantis system}
   
\titlerunning{Constraining the orbit of the retrograde planet in the $\nu$ Octantis system}
 \authorrunning{Signor et al.}

   \author{Alan Cefali Signor
          \inst{1,2},
          Alexandre C. M. Correia\inst{1,3}, Maria Helena Moreira Morais \inst{2},\\
          Timothée Vaillant \inst{1}, Ho Wan Cheng \inst{4}, Man Hoi Lee \inst{4,5,6} and Trifon Trifonov \inst{7,8}}
   \institute{CFisUC, Departamento de Física, Universidade de Coimbra, 3004-516 Coimbra, Portugal
  % \\ \email{acor@uc.pt}
         \and
             IGCE, Universidade Estadual Paulista (UNESP), Av. 24-A, 1515, 13506-900 Rio Claro, SP, Brazil 
          % \\  \email{alan.cefali@unesp.br; helena.morais@unesp.br}
          \and 
             LTE, Observatoire de Paris, Universit\'e PSL, Sorbonne Universit\'e, CNRS, 75014 Paris, France
          \and 
             Department of Earth and Planetary Sciences, The University of Hong Kong, Pok Fu Lam, Hong Kong
          \and 
             Department of Physics, The University of Hong Kong, Pok Fu Lam, Hong Kong
          \and
             Hong Kong Institute for Astronomy and Astrophysics, The University of Hong Kong, Pok Fu Lam, Hong Kong
          \and 
             Zentrum für Astronomie der Universität Heidelberg, Heidelberg, Germany
          \and 
             Department of Astronomy, Faculty of Physics, Sofia University St. Kliment Ohridski, Sofia, Bulgaria}
\date{\today; Received; accepted To be inserted later}

 \abstract{
The $\nu$ Octantis system is composed of a compact stellar binary and a retrograde S-type planet orbiting near the edge of stability. Here, we investigate the dynamically viable architectures of this system by exploring the planetary orbital elements with a Monte Carlo method. To handle the large number of free parameters, we employed a novel approach in which all relevant orbital parameters were varied simultaneously, which we refer to as corner stability maps. Furthermore, since the osculating orbital period of the planet undergoes large oscillations, we combined these maps with frequency analysis to identify the stable configurations that remain consistent with the observational data. We find that the retrograde planetary orbit most likely lies in the region of influence of the high-order 28/$-$11 mean-motion resonance, where its long-term stability is enhanced by secular apsidal alignment.}


   \maketitle

\section{Introduction}

The $\nu$~Octantis system comprises a primary K1~IV subgiant star, $\nu$-Oct~A, with mass $m_0 = 1.57$~M$_\odot$, and a secondary star, $\nu$-Oct~B, with mass $m_2 = 0.57$~M$_\odot$, now identified as a white dwarf. The binary orbit has a semi-major axis $a_2 = 2.61$~au and an eccentricity $e_2 = 0.237$ \citep{ramm2016conjectured, cheng2025}.
Using astrometric and spectroscopic analysis of the system, \citet{ramm2009spectroscopic} detected a low-amplitude, 418-day periodic signal in the radial-velocity (RV) residuals of $\nu$-Oct~A. 
These variations could be due to rotational modulation of surface phenomena, stellar pulsations, or the perturbation of a planetary mass companion \citep{ramm2009spectroscopic}.
\citet{ramm2015line} and \citet{ramm2021photospheric} explored the first two explanations in detail and showed that $\nu$-Oct~A is a quiet star.
As a consequence, they concluded that the most likely explanation for the RV periodic oscillation is the presence of a Jupiter-mass companion orbiting $\nu$-Oct~A in an S-type configuration\footnote{In an S-type configuration the planet orbits around a single component of the binary \citep{dvorak1984numerical}.}. \cite{morais2012precession} proposed an alternative dynamical mechanism to explain the 418-day signal.
If the secondary star $\nu$-Oct~B were a double star system, it could mimic a signal that is induced by a planet around $\nu$-Oct~A.
At the time, the amount and precision of the observational data were insufficient for researchers to distinguish between the two scenarios. \citet{ramm2016conjectured} later showed that the binary orbit does not have the necessary apsidal motion for this alternative explanation. 

However, the planetary hypothesis was also highly questionable.
Given that the secondary star is only $1.99$~au away at pericentre, the presence of the then-hypothetical planet on a near-circular orbit around the primary star with a semi-major axis $a_1 = 1.24$~au would pose serious stability concerns. According to the stability boundary criterion obtained for coplanar prograde orbits in S-type configurations by \citet{holman1999long}, a planet around $\nu$-Oct~A can only be stable for $a_1 < 0.65$~au.

\citet{gayon2009fitting} showed that for some systems the RV fits have smaller residuals for retrograde configurations than for prograde configurations.
Moreover, planets on retrograde orbits can be more stable than the prograde orbits \citep[e.g.][]{morais2012stability, giuppone2017, quarles2016}.
A retrograde version of the stability boundary equation was obtained by \citet{lee2024}. Considering the masses and eccentricity of the $\nu$-Oct system, the stability limit is comparable to the planet’s semi-major axis determined from the observations.

Several studies have explored the possibility of a retrograde planetary orbit in the $\nu$-Oct system.
\citet{eberle2010reality} investigated both prograde and retrograde cases and showed that only a retrograde configuration has long-term orbital stability.
\citet{quarles2012stability} and \citet{gozdziewski2013} further supported this scenario by considering a wider range of initial conditions and observational uncertainties, finding that the proposed planet is indeed only possible on a retrograde orbit. 
Since the orbital period of the binary, $P_2 = 1051$~day, was close to a $5/2$ ratio with the periodic signal at 418-day \citep{ramm2009spectroscopic}, \citet{gozdziewski2013} investigated whether the suggested planet could be trapped in a mean motion resonance (MMR) and identified the stable MMR regions as a function of the planet's semi-major axis and mutual inclination.
They showed that the retrograde 5/$-$2~MMR can be stable, although their best-fit orbital solutions were confined to tiny stable regions in the middle of a mostly chaotic phase space.

Recently, \citet{cheng2025} confirmed the existence of a retrograde planet in $\nu$ Octantis by presenting stable orbital solutions that are consistent with the RV data and plausible planet-formation scenarios.
The recent identification of $\nu$-Oct~B as a white dwarf led \citet{cheng2025} to show that standard first-generation in situ formation around $\nu$-Oct~A is particularly problematic. 
One possible scenario is that the planet formed on a circumbinary orbit and was later transferred to a retrograde circumprimary orbit through planet--planet scattering followed by capture, a mechanism shown to produce S-type planets in close binaries \citep{gong2018, cheng2025}. 
Another possibility is a second-generation origin, in which mass lost by the progenitor of the present white dwarf formed an accretion disc around $\nu$-Oct~A, providing fresh material for planet formation after the main episode of binary evolution \citep{perets2013, cheng2025}. 
More generally, strong external perturbations such as stellar fly-bys or exchange encounters could also implant a planet formed in a less hostile environment onto a highly inclined or retrograde orbit \citep{malmberg2011effects, breslau2019}. While the origin of the planet in $\nu$~Octantis remains an open question, all plausible pathways require substantial post-formation dynamical reconfiguration and/or binary evolution.

Previous works such as \citet{eberle2010reality}, \citet{quarles2012stability}, and \citet{gozdziewski2013} explored the stability of the $\nu$-Oct system by considering the planet and the secondary at the apoapsis of their orbits and fixing their initial eccentricity and longitudes, while \cite{cheng2025} focused on the subset of configurations favoured by the RV posterior distribution. 
Therefore, a dynamical analysis of the planet's long-term stability, considering a wider set of positions, orientations, and eccentricities, has yet to be carried out. 
This is not a minor detail because the mutual perturbations in the $\nu$~Oct system are very strong, and all the osculating orbital elements of the planet are therefore expected to undergo large-amplitude variations.

In the present study, we determine the possible stable configurations of the retrograde planet in the $\nu$~Oct binary system. In Sect.~\ref{sec2}, we revisit the RV observational constraints and use them to delimit the orbital configurations explored in our dynamical analysis. In Sect.~\ref{sec3}, we assume a coplanar and retrograde orbit for this planet and determine the combinations of orbital parameters that are stable and best fit the observations. In Sect.~\ref{sec4}, we relax the constraint on coplanar retrograde orbits and explore the full space of parameters. Finally, we discuss our results in Sect.~\ref{sec5}.

\section{Preliminary orbital solutions}
\label{sec2}

The $\nu$-Oct system has been observed on several occasions using spectroscopic measurements.
The High Efficiency and Resolution Canterbury University Large Échelle Spectrograph (HERCULES) mounted on the 1-m McLellan telescope at the Mt. John University Observatory (New Zealand) was used to collect a large set of data between 2001 and 2013.
Using different methods with distinct precisions, a total of 1\,437~RVs of $\nu$-Oct~A were obtained \citep{ramm2009spectroscopic, ramm2015line, ramm2016conjectured}.
These measurements are provided in Tables~3, 4, and 5 of \citet{ramm2016conjectured}, with mean precisions of 3.6, 18.7 and 15.0~m/s, respectively.
From 2018 to 2019, the system was observed using the High Accuracy Radial velocity Planet Searcher (HARPS) spectrograph, mounted on the European Southern Observatory (ESO) 3.6-m telescope at La Silla (Chile) under programme IDs: 0100.C-0414, 0101.C-0232, and 0103.C-0548 \citep{cheng2025}. 
These data consist of 213~RVs of $\nu$-Oct~A with a mean precision of 2.9~m/s.

In this work we use all 1\,650~RVs from both HERCULES and HARPS spectrographs spanning $\sim 18$~years of observations.
Assuming a three-body model, we obtained orbital solutions for the $\nu$-Oct system adopting the Jacobi reference frame for the orbital elements, where $a$ is the semi-major axis, $e$ is the eccentricity, $I$ is the inclination to the line of sight, $\omega$ is the argument of pericentre, $\Omega$ is the longitude of the ascending node, $\lambda$ is the mean longitude, $P$ is the orbital period, and $K$ is the semi-amplitude of the radial-velocity variations. 
The subscripts 1 and 2 refer to the planet and to $\nu$-Oct~B, respectively.

\subsection{Keplerian fit}

\begin{table}
\centering
\caption{Orbital parameters for the two bodies orbiting $\nu$-Oct~A, with $m_0 = 1.57$~M$_\odot$, obtained
with a Keplerian fit.} 
\label{Table 1}
\renewcommand{\arraystretch}{1.3}
\begin{tabular}{l @{\hspace{1.3cm}}c @{\hspace{0.8cm}} c}
\toprule
\toprule
\multicolumn{3}{c}{{\bf Keplerian Best Fit} ($\pm 1 \sigma$)} \\
\midrule
{\bf Parameter} & $\boldsymbol{\nu}$ {\bf Oct Ab} &  $\boldsymbol{\nu}$ {\bf Oct B} \\
\midrule
$P$ [day] & 411.80\(^{+0.38}_{-1.16}\) & 1050.013\(^{+0.002}_{-0.060}\)\\
$K$ [m/s] & 41.50\(^{+2.34}_{-1.06}\) & 7049.99\(^{+2.67}_{-1.95}\)  \\
$\lambda$ [deg] &  175.69\(^{+2.014}_{-9.00}\) &   140.822  \(^{+0.007}_{-0.075}\) \\
$e$ & 0.0343\(^{+0.0741}_{-0.0132}\)  & 0.2363\(^{+0.0003}_{-0.0002}\) \\
$\omega$ [deg] & 133.94\(^{+141.76}_{-51.69}\) & 75.19\(^{+0.04}_{-0.10}\)  \\
\midrule
$m \, \sin(I)$& 2.10\(^{+0.08}_{-0.12}\) $M_\mathrm{Jup}$ & 0.5503\(^{+0.0001}_{-0.0249}\) $M_\odot$ \\
$a$ [AU] & 1.2755\(^{+0.0008}_{-0.0333}\) & 2.6227\(^{+0.0003}_{-0.0599}\) \\
\midrule
rms [m/s] &\multicolumn{2}{c}{6.24} \\
$\lnL$ & \multicolumn{2}{c}{-1\,226.25}\\
\bottomrule
\end{tabular}
\tablefoot{The fit uses the date of the first observation as the reference date (JD~2\,452\,068.0607).}
\end{table}

\begin{figure}
    \centering
    \includegraphics[scale=0.68]{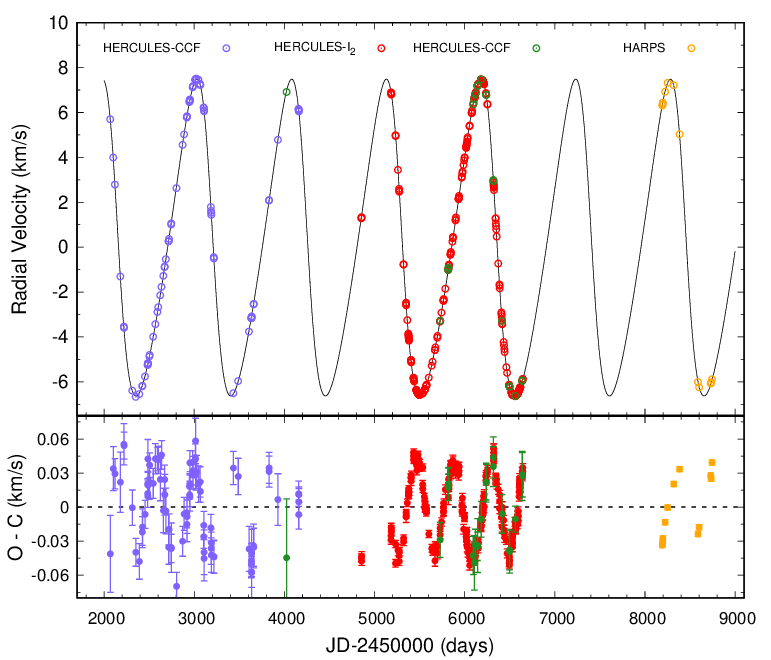}
    \caption{Radial  velocities of $\nu$-Oct obtained with HERCULES and HARPS, superimposed on a single Keplerian orbital solution (\textit{top}) and the residuals (\textit{bottom}).}
    \label{fig:1}
\end{figure}

We first fitted the complete set of RVs of $\nu$-Oct~A with a single orbiting companion; that is, we considered a two-body system on a Keplerian orbit (Table~\ref{Table 1}).
In Fig.~\ref{fig:1} (top) we plot the RVs superimposed on a single companion model, which shows good agreement with the observations.
However, in Fig.~\ref{fig:1} (bottom), we additionally plot the residuals of this model and observe some periodic oscillation with an amplitude of about 60~m/s.

The periodic nature of these residuals, with an estimated period of $\sim 400$~days is already present in the early RV measurements obtained with HERCULES and interpreted by \citet{ramm2009spectroscopic} as evidence of a possible planet.
We note that this oscillation is still present in the HARPS data, although these data only cover one oscillation cycle.
However, since the HARPS data were taken several years after the last HERCULES point, they are very useful for checking whether the signal is still coherent \citep{cheng2025}.

\begin{figure}
    \centering
    \includegraphics[scale=0.65]{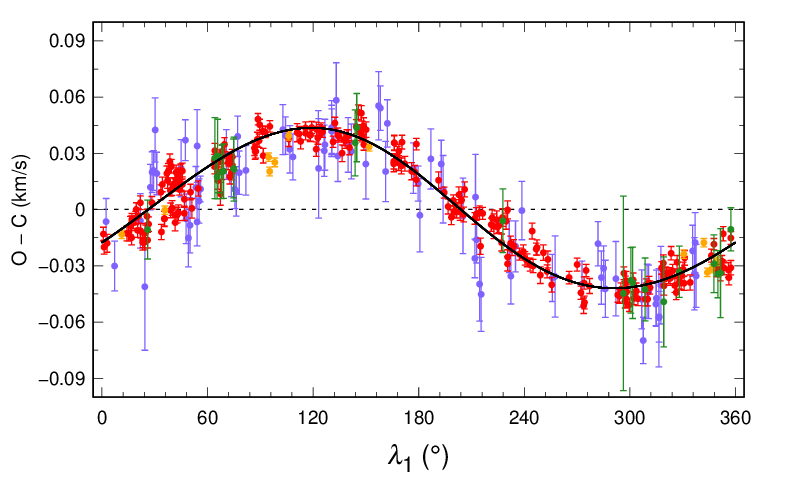}
    \caption{Phase-folded RV curves for the planet in the $\nu$-Oct system using a two-Keplerian model. The contribution of the $\nu$-Oct~B signal (Fig.~\ref{fig:1}) was subtracted.}
    \label{fig:vrphase}
\end{figure}

We then fit the complete RVs dataset using the same RV modelling scheme as in \citet{cheng2025},
using the {\sc Exo-Striker} toolbox\footnote{\url{https://github.com/3fon3fonov/exostriker}} \citep{Trifonov2019es}. We applied a 
double Keplerian model, simultaneously fitting the Doppler signals induced by both the stellar and planetary companions. Our 
modelling also optimises the individual RV dataset offsets and jitter terms, the latter being added in quadrature to the formal 
uncertainties to account for stochastic RV noise \citep{Baluev2009}. In addition, we adopted a Gaussian process rotational kernel 
\citep[see,][]{celerite} to model the RV signal induced by stellar activity, following \citet{cheng2025}. To efficiently explore the 
high-dimensional parameter space, we employed a dynamic nested sampling scheme \citep{Skilling2004} via the {\tt dynesty} sampler 
\citep{Speagle2020}, followed by a Bayesian posterior analysis of the model parameters. We adopted 100 live points per fitted parameter 
and used the random-walk sampling option to ensure robust posterior convergence. From the posteriors, we extracted the sample with 
the maximum \(\ln\mathcal{L}\), i.e. the best-fit solution, and the 1$\sigma$ posterior uncertainties. 
From the maximum, we obtain periods $P_1$ = 411.80$^{+0.38}_{-1.16}$ days and $P_2$ = 1050.013$^{+0.002}_{-0.060}$ days, and 
minimum masses of $m_1 \sin(I) = 2.10^{+0.08}_{-0.12}$~M$_\mathrm{Jup}$ and $m_2 \sin(I) = 0.5503^{+0.0001}_{-0.0249} M_\odot$ for $\nu$-Oct~Ab and $\nu$-Oct~B, respectively. Table~\ref{Table 1} lists the parameters of the best fit and the posterior mean estimates of this Keplerian fit.
Figure~\ref{fig:vrphase} shows the phase-folded RV measurements after subtracting the $\nu$-Oct~B signal, superimposed on the fitted orbit of the Jupiter-mass companion.
We observe good agreement in terms of amplitude and period, but more importantly, we verify that the HARPS data are still in phase (although they do not cover the whole curve), which reinforces the presence of a planetary companion in the system.

\subsection{Newtonian fit}

The inclination with respect to the plane of the sky, $I$, and the longitude of the node, $\Omega$, are not constrained by a Keplerian fit, and hence we only obtain minimum estimates for the masses (Table~\ref{Table 1}).
However, astrometric measurements of the $\nu$-Oct system taken by \textsc{Hipparcos} can help to determine some of the missing parameters.
Indeed, \citet{ramm2009spectroscopic} obtained $ I_2 = ( 70.8 \pm 0.9 )^\circ$ and $\Omega_2 = ( 87 \pm 1 )^\circ $ for $\nu$-Oct~B, by combining \textsc{Hipparcos} astrometric data with HERCULES spectroscopic measurements, which yields a mass $m_2 = 0.585 \pm 0.003 $~M$_\odot$.

\citet{cheng2025} adopted $I_2 = 71.8^\circ$ and $\Omega_2 = 86.5^\circ$ to model the RV data. These values were also obtained by fitting the $\nu$-Oct binary with the \textsc{Hipparcos} astrometric data. The resolution of the \textsc{Hipparcos} data is insufficient to constrain the orbital parameters of the planet.  \citet{cheng2025} also determined the mass of the primary star $m_0 = 1.57 \pm 0.06$~M$_\odot$ using a Bayesian inference scheme.

To test the stability of the system, for simplicity, it is often assumed that the orbits are prograde and coplanar, that is, $I_1=I_2$ and $\Omega_1 = \Omega_2$.
With this assumption, the system provided in Table~\ref{Table 1} is totally unstable and the planet is ejected just after only a few orbital revolutions.
Previous studies on the stability of the planet in the $\nu$-Oct system have already reported on this problem, but have concluded that the system can be stable if the planet is on a retrograde orbit \citep{eberle2010reality, quarles2012stability, gozdziewski2013, ramm2016conjectured, lee2024, cheng2025}.
A retrograde coplanar orbit corresponds to a mutual inclination $i = 180^\circ$ with respect to the orbital plane of binary stars, which can be converted into the plane of the sky as $I_1 = 108.2^\circ$ and $\Omega_1 = 266.5^\circ$ (Eqs.\,\ref{eq1} and \ref{eq4}).
However, when we adopt these values, the system provided in Table~\ref{Table 1} still becomes unstable rapidly.

The orbital solution obtained with a Keplerian fit (Table~\ref{Table 1}) corresponds to mean orbital elements that do not change with time (except for the mean longitudes, $\lambda$).
When mutual interactions between all bodies are significant, as appears to be the case for the $\nu$-Oct system, the osculating orbital parameters at the reference date can show large deviations from the mean Keplerian elements.
Therefore, to take into account the effect of these mutual perturbations, we considered the three-body Newtonian fit obtained by \citet{cheng2025}. This fit assumes a retrograde coplanar planetary orbit and is the preferable stable fit from the posterior samples (Table~\ref{Table 2}).

\begin{table}
\centering
\caption{Orbital parameters for the two bodies orbiting $\nu$-Oct~A, with $m_0 = 1.57$~M$_\odot$, obtained
from a three-body Newtonian fit.}
\label{Table 2}
\renewcommand{\arraystretch}{1.3}
\begin{tabular}{l @{\hspace{1.3cm}}c @{\hspace{0.8cm}} c}
\toprule
\toprule
 \multicolumn{3}{c}{{\bf Preferable Stable Fit}  ($\pm 1 \sigma$)} \\
\midrule
{\bf Parameter} & $\boldsymbol{\nu}$ {\bf Oct Ab} & {\bf $\boldsymbol{\nu}$ Oct B } \\
\midrule
$P$ [day] & 402.36\(^{+7.66}_{-6.00}\) & 1050.74\(^{+0.20}_{-0.10}\) \\
$K$ [m/s] & 42.74\(^{+2.58}_{-1.58}\) & 7046.77\(^{+4.95}_{-0.88}\) \\
$\lambda$ [deg]  & 177.44$^{+7.53}_{-4.06}$ & 140.71$^{+0.02}_{-0.11}$ \\
$e$  & 
0.1953\(^{+0.0498}_{-0.0367}\) & 0.2366\(^{+0.0003}_{-0.0003}\) \\
$\omega$ [deg]  & 
100.43\(^{+32.49}_{-15.20}\) & 74.88\(^{+0.06}_{-0.14}\) \\
$I$ [deg]  & 
108.20 & 71.80\(^{+0.70}_{-0.60}\) \\
$\Omega$ [deg]  & 
266.50 & 86.50\(^{+0.30}_{-0.30}\) \\
\midrule
Mass & 2.17\(^{+0.14}_{-0.09}\)  $M_\mathrm{Jup}$ & 0.5727\(^{+0.0127}_{-0.0134}\) $M_\odot$ \\
$a$ [AU] & 1.2445\(^{+0.0184}_{-0.0239}\) & 2.6087\(^{+0.0292}_{-0.0302}\) \\
\midrule
rms [m/s] &\multicolumn{2}{c}{5.97} \\
$\lnL$ & \multicolumn{2}{c}{-1\,230.3}\\
\bottomrule
\end{tabular}
\tablefoot{The fit uses the reference date of the first observation (JD~2\,452\,068.0607).}
\end{table}

\begin{figure}
\centering
    \includegraphics[scale=0.13]{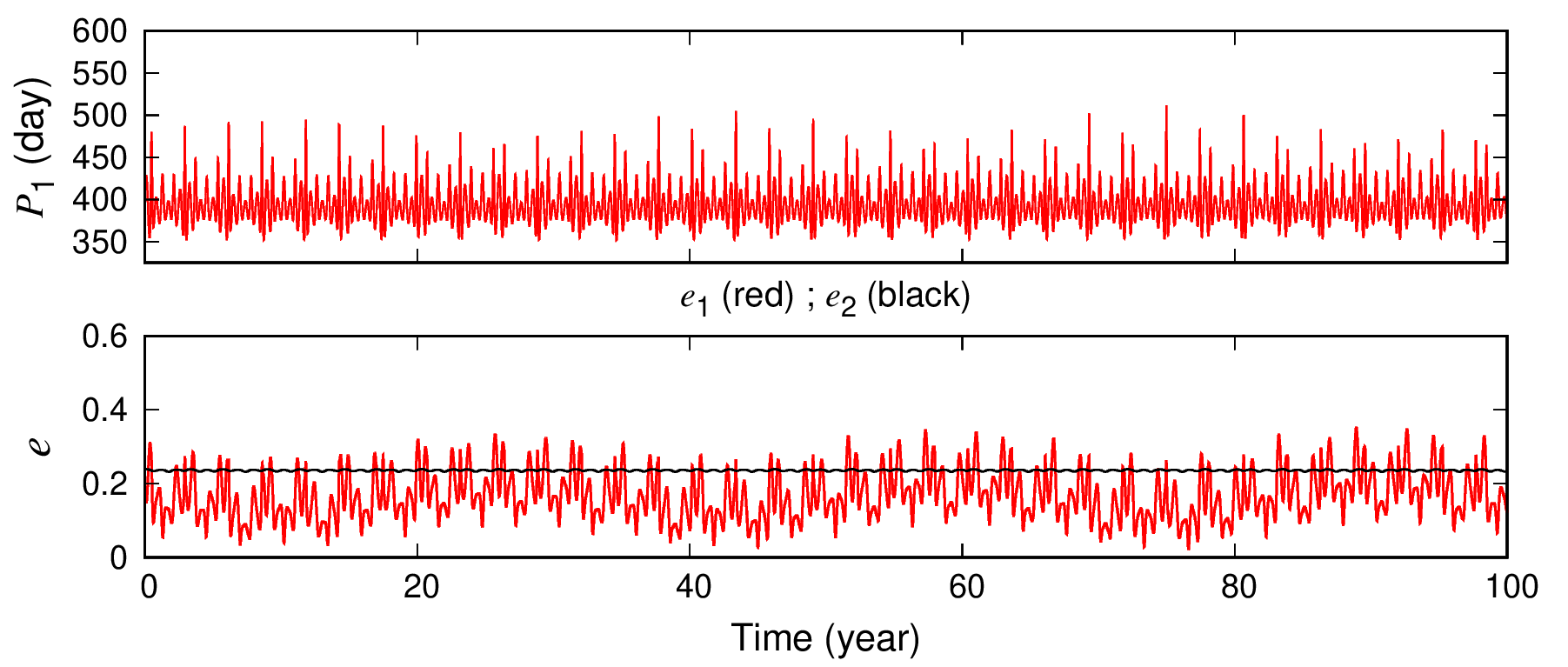} \\
      \includegraphics[scale=0.13]{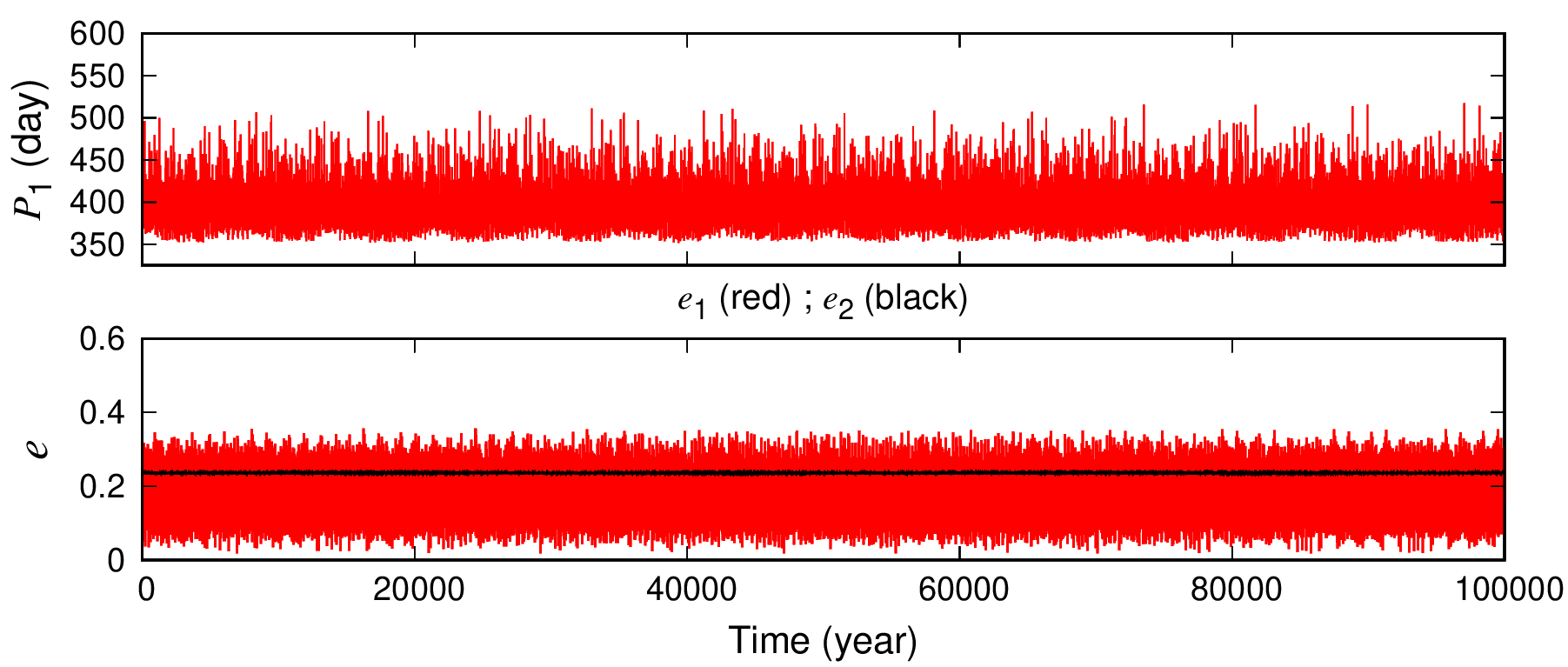}
    \caption{Orbital evolution of the $\nu$-Oct system for 100~yr (\textit{top}) and 100~kyr (\textit{bottom}) using the initial condition of the best-fit three-body Newtonian model (Table~\ref{Table 2}). We show the orbital period of the planet and the eccentricities of the planet and $\nu$-Oct~B.
\label{fig:2}}
\end{figure}

Figure~\ref{fig:2} shows the evolution of the orbital period and the eccentricity of the planet over 100~yr and 100~kyr obtained with the Newtonian fit (Table~\ref{Table 2}).
We observe that the system is much more stable than that obtained with the Keplerian fit (Table~\ref{Table 1}).
Moreover, we confirm that the orbital parameters exhibit large oscillations owing to the strong mutual perturbations.
The osculating orbital period ranges from 352 to 517 days, while the eccentricity varies from 0.018 to 0.36.
These large oscillations explain the substantial differences between the osculating orbital elements of the Newtonian fit (Table~\ref{Table 2}) and the mean orbital elements obtained with the Keplerian fit (Table~\ref{Table 1}). Therefore, to achieve a clearer understanding of orbital solutions that are simultaneously compatible with current RV observational data and stable over long timescales ($\sim$~Gyr), we performed a detailed dynamical analysis of the $\nu$-Oct system.

\section{Retrograde coplanar orbits}
\label{sec3}

We assumed that the orbital parameters of $\nu$-Oct~B are well determined and used the reference values provided in Table~\ref{Table 2}. 
For simplicity, we first studied the dynamics of the $\nu$-Oct system presuming that the planet is on a retrograde coplanar orbit, i.e. we fixed $I_1 = 108.2^\circ$ and $\Omega_1 = 266.5^\circ$ as in Table~\ref{Table 2}.
For the masses, we adopted $m_1 = 2.17 $~M$_\mathrm{Jup}$ and $m_2 = 0.5723$~M$_\odot$ (Table~\ref{Table 2}). 

This left four orbital parameters of the planet to constrain: the orbital period, $P_1$, the mean longitude of the date, $\lambda_1$, the eccentricity, $e_1$, and the argument of pericentre, $\omega_1$.
These parameters are the most uncertain to extract from RV surveys (Table~\ref{Table 2}), but more importantly, in the $\nu$-Oct system their osculating values undergo large variations in time (Fig.~\ref{fig:2}). 
Given the observed variations in these parameters, we generated $6 \times 10^6$ different initial conditions using random uniform distributions spanning the following intervals: $P_1 \in [288,580]$~day, $\lambda_1 \in [0^\circ,360^\circ]$, $e_1 \in [0,0.5]$, and $\omega_1 \in [0^\circ,360^\circ]$.

\subsection{Stability analysis}

We numerically integrated the $\nu$-Oct system using REBOUND with the adaptive time-step Bulirsch-Stoer integrator \citep{rein2012rebound} and the precision set at $10^{-12}$. We integrated each initial condition for 12~kyr ($\approx 4174 \, P_2$) with an output step of 0.3~yr.
We stopped the integration when the distance between two bodies was smaller than the sum of their radii or when the distance to $\nu$-Oct~A was greater than 5 au. 

To quickly inspect the stability of each trajectory, we adopt the frequency analysis method \citep{laskar1993frequency} to map the diffusion of the orbital evolution.
Using the computer algebra system TRIP \citep{TRIP}, we performed a frequency analysis of the mean longitude of the planet, $\lambda_1$, over successive time intervals $[0,6]$~kyr and $[6,12]$~kyr, and determined the main frequency in each interval, $n_1$ and $n_1'$, respectively.
The stability of the orbit is measured by the index
\begin{equation}
\DeltaIn \equiv \left\lvert1 - {n_1'}/{n_1}\right\rvert \ ,
\label{diffusionindex}
\end{equation}
which estimates the stability of the long-term diffusion of the mean motion \citep{dumas1993}.

For regular and stable orbits, the diffusion index tends to zero, but takes high values for chaotic motion.
It is difficult to estimate what is the value of $\DeltaIn$ for the stability threshold that allows us to distinguish between these two cases.
This value depends on the dynamical system and the time interval considered.
To estimate the stability threshold, we applied the procedure proposed by \cite{couetdic2010dynamical} to the $\nu$-Oct system and find that after 12~kyr most of the solutions obtained with $\DeltaIn < \DeltaTrans$ can be considered stable over this time interval. We stress that this integration time is used as a fast diffusion diagnostic for large Monte Carlo samples, not as a stand-alone proof of Gyr stability.

\subsubsection{Corner stability maps}
\label{CSMap}

We explored four orbital parameters in our simulations, but for clarity reasons, it is desirable to show the results only in 2D plots.
As in many multi-parameter orbital fitting problems, we can use corner plots, which consist of a set of 2D combinations of all free parameters; that is, an $N$-dimensional parameter space can be represented by $N_{2D} = N \,(N-1)/2$ two-dimensional graphs. 
The standard analysis of corner plots is based on the projection of all initial conditions in 2D phase spaces, identifying the regions with the largest number of points and representing them using contour plots \citep[e.g.][]{foreman2017corner}.
In this work, we did not simply count the number of stable trajectories, because each has a different stability index (Eq.\,\ref{diffusionindex}).
Therefore, we introduced a new method that combines stability maps with corner plots, which we call corner stability maps. 
There are three main steps to the construction of these maps:

\begin{enumerate}
\item Group the random initial conditions onto an equally spaced grid, such that each pixel of the 2D graph gathers information from a different set of initial conditions;
\item Choose a stability index that best represents the majority of the initial conditions present in each pixel;
\item Adopt a metric to display the stability index that best characterises each pixel.
\end{enumerate}

Each pixel should contain a minimum number of points in order to perform a statistical selection of the best representative index.
Since we integrated $6 \times 10^6$ initial conditions, we adopted a $200 \times 200$ pixel mesh, such that each pixel contains, on average, information from 150 simulations.
For instance, to build the ($\omega_1,e_1$) map, each pixel has a size of ($1.8^\circ \times 0.0025$) and contains all the initial conditions whose $e_1$ and $\omega_1$ differ only by that interval size.

After several trials, we find that the median value of the diffusion indices of all initial conditions grouped in a pixel best represents the stability index of the whole pixel.
The mode and mean also provide similar results, but the median is more balanced between the peak of the distribution (mode) and its spread (mean).
The median also provides better contrasts between the different indices in the maps. 

Finally, to avoid the difference between the weights of low and high values of the diffusion indices when computing the median, we represent the diffusion by $\log_{10} \DeltaIn$.
That is, stable trajectories are given by $\log_{10} \DeltaIn < - 5$.
The diffusion index is represented by a calibrated colour scale such that blue and cyan correspond to stable trajectories, while orange and red correspond to chaotic motion.

The corner stability maps are not intended to replace the posterior sampling of the observational model. Instead, the posterior sub-samples define the observationally allowed region, while the maps provide the local dynamical context by showing how this region is embedded in the surrounding low-diffusion secular and resonant structures. We describe the connection between these two sources of information in more detail in the following subsections.

\subsubsection{Filtering the results}

The osculating orbital parameters ($P_1, \lambda_1, e_1, \omega_1$) undergo large variations, but the mean orbital period and the planet's phase angle need to be relatively well determined, since we expect quasi-periodic motion (we assume that the system is stable over long timescales, as shown in \citealt{cheng2025}). 
Interestingly, these two parameters can be directly obtained from the observational data using the Keplerian fit, which provides a well-defined sinusoidal signal (Fig.~\ref{fig:vrphase}).
As a result, we imposed additional observational constraints $\Pm = 411.8$~day and $ \Lm = 175.69^\circ$, where $\Lm$ is the value of $\lambda_1$ for the reference date of the first observation, JD 2\,452\,068.0607 (Table~\ref{Table 1}). 

Since we used the frequency analysis of the mean longitude, $\lambda_1$, to determine the diffusion index, we can also extracted the mean orbital periods as $\Pm = 2 \pi / n_1$ and $\Pms = 2 \pi / n_2$, where $n_1$ and $n_2$ are the main frequencies of $\lambda_1$ and $\lambda_2$, respectively.
As an example, in Table~\ref{tab:freq_planet} we show the quasi-periodic decomposition of $\lambda_1$ after performing the frequency analysis of the Newtonian best stable fit (Table~\ref{Table 2}).
We observe that the main frequency, $n_1 = -0.8722596783^\circ$/day, corresponds to a period $\left|2\pi/n_1\right| = 412.7211$~days, which is relatively close to the Keplerian value (Table~\ref{Table 1}). The mean motion of the planet, $n_1$, is negative because of its retrograde orbit.

\begin{figure}
\centering
  \includegraphics[width=0.94\columnwidth]{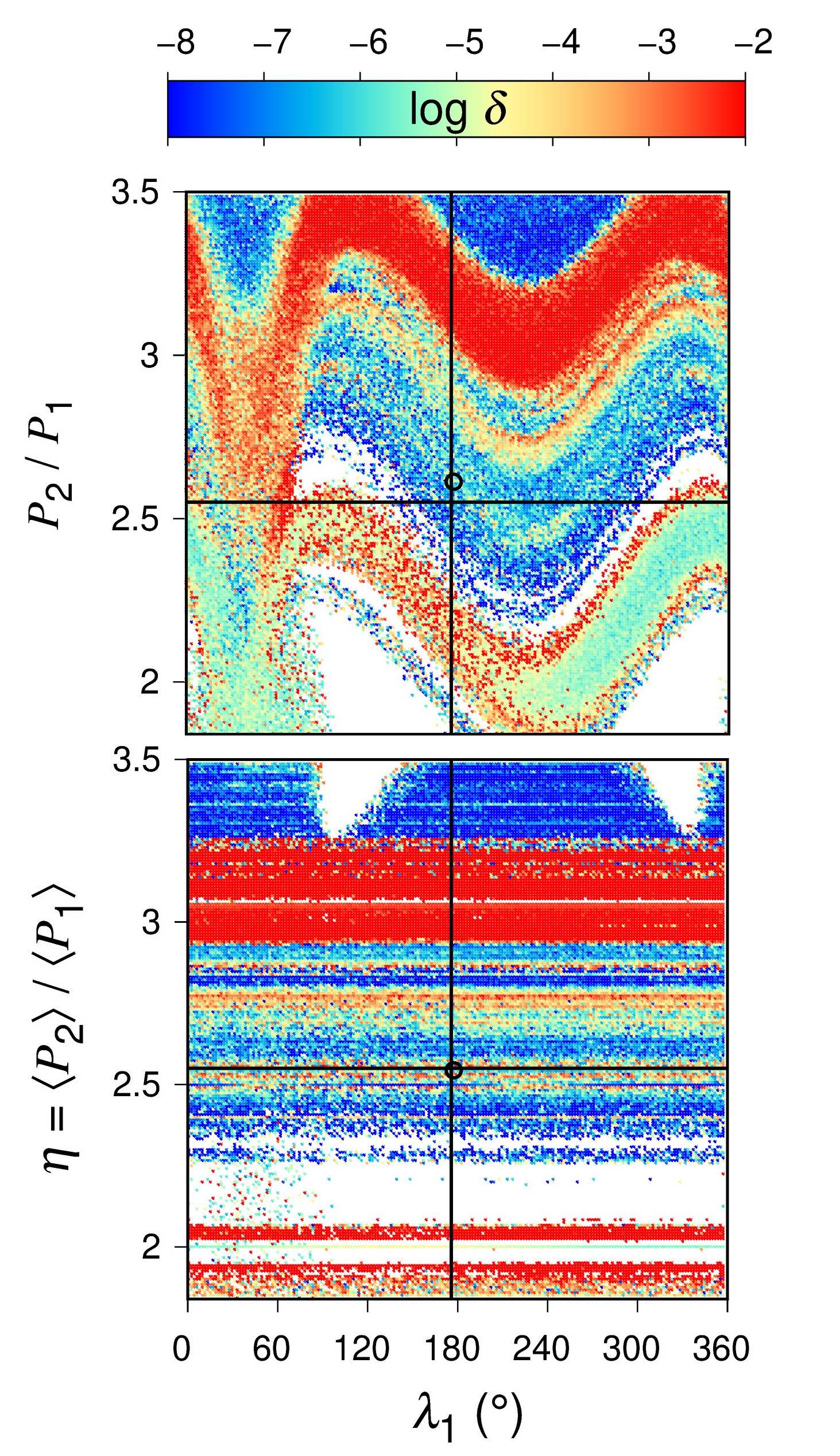}
  \caption{Corner stability map for ($\lambda_1, P_2/P_1$) (\textit{top}) and ($\lambda_1, \eta$) (\textit{bottom}). The straight lines indicate the $P_2/P_1$ and $\lambda_1$ values corresponding to the Keplerian fit (Table~\ref{Table 1}), while the circle corresponds to the Newtonian best-fit value (Table~\ref{Table 2}). The colour scale corresponds to the diffusion index in logarithmic scale (Eq.~\ref{diffusionindex}).}
  \label{ratiocomparison}
\end{figure}

Figure \ref{ratiocomparison} presents a comparison between the ($\lambda_1, P_2/P_1$) and the ($\lambda_1,\eta$) stability maps, with
\begin{equation}
\eta \equiv \Pms/\Pm = \left|n_1/n_2 \right|,
\end{equation}
that is, one map considering the initial osculating orbital periods and another map considering the corresponding mean orbital periods (filtered map).
For guidance, we also plot the $\lambda_1$ and $P_2/P_1$ values corresponding to the Keplerian fit (Table~\ref{Table 1}) as straight lines and the value corresponding to the Newtonian fit (Table~\ref{Table 2}) with a circle.
We observe that the mean orbital periods of the Newtonian fit correspond approximately to the orbital periods of the Keplerian fit. 
We chose to represent $\eta = \Pms/\Pm$ instead of $\Pm$ because $P_2$ also undergoes some variations due to the planet, which allowed us to easily identify MMRs as straight lines (see Sect.~\ref{coplanresults}).

In Fig.~\ref{ratiocomparison} we observe that the corner stability map, in which we varied four parameters at a time, is able to capture the essence of the classic 2D stability maps, in which we varied only two orbital parameters and kept the remaining parameters constant.
We identify white regions corresponding to trajectories that collided with or were ejected from the system within 12~kyr, that is, initial conditions that did not survive the integration time because of collision or ejection.
We also observe red regions of strong chaotic diffusion alternating with blue regions corresponding to stable orbits.
While in the white regions no solution survives, in the red regions we cannot exclude some stable solutions, and in the blue regions we cannot exclude some unstable solutions.
However, the fact that some regions are clearly defined means that most of the trajectories in the area behave the same way.

Figure~\ref{ratiocomparison} also shows that the stability map is much more regular in the filtered case and thus easier to compare with the mean best fit solution (given by the intersection of the two straight lines).
Indeed, while the ($\lambda_1, P_2/P_1$) map provides almost no observational constraint on $P_1$ owing to its large variations, the ($\lambda_1, \eta$) map provides a strong restriction on $\Pm$, which allowed us to identify which initial conditions are more reliable representations of the real system.
As a result, in the remainder of this work, we construct the corner stability maps using the mean orbital periods ratio, $\eta = \Pms/\Pm = \left|n_1/n_2 \right|$. Thus, the horizontal dashed lines in the following maps and the corresponding resonant labels refer to commensurabilities between the proper frequencies. This convention avoids identifying an MMR from the instantaneous value of $P_2/P_1$, which can be misleading in such a strongly perturbed system.

\subsubsection{Confidence levels}
\label{sec:confidence}

The Newtonian best stable fit (Table~\ref{Table 2}), which
has the highest $\lnL$, provides the current best representation of the system, but not necessarily the correct one. Any other orbital solution with similar $\lnL$ values is also a reliable description of the system. In general, we expect that the trustworthiness of a given solution increases as its $\lnL$ value approaches the maximum value. As a result, for neighbouring initial conditions, $\lnL$ is not expected to change much.
Therefore, in traditional 2D stability maps, where we vary only two orbital parameters at a time and each pixel contains information pertaining to a single orbital solution, we often plot $\lnL$ (or similarly ${\chi^2}$) contour level curves that show the likelihood of a given region of the phase space \citep[e.g.][]{correia2005coralie, correia2009harps}.

With corner stability maps, such level curves cannot be defined, since each pixel combines information from many different initial conditions, each with its own $\lnL$ adjustment. As a consequence, we adopted a different approach to determine the region of phase space, which is in agreement with the observations. For instance, one might adopt the maximum $\lnL$ of all orbital solutions within a pixel, but this solution may correspond to a special case that does not translate the fitness of most solutions in that pixel. Another issue is that this approach does not ensure a smooth transition between the $\lnL$ values of neighbouring pixels. Indeed, after testing this method, we observed that the level curves were not correctly defined.

\begin{figure}
\centering
  \includegraphics[width=0.95\columnwidth]{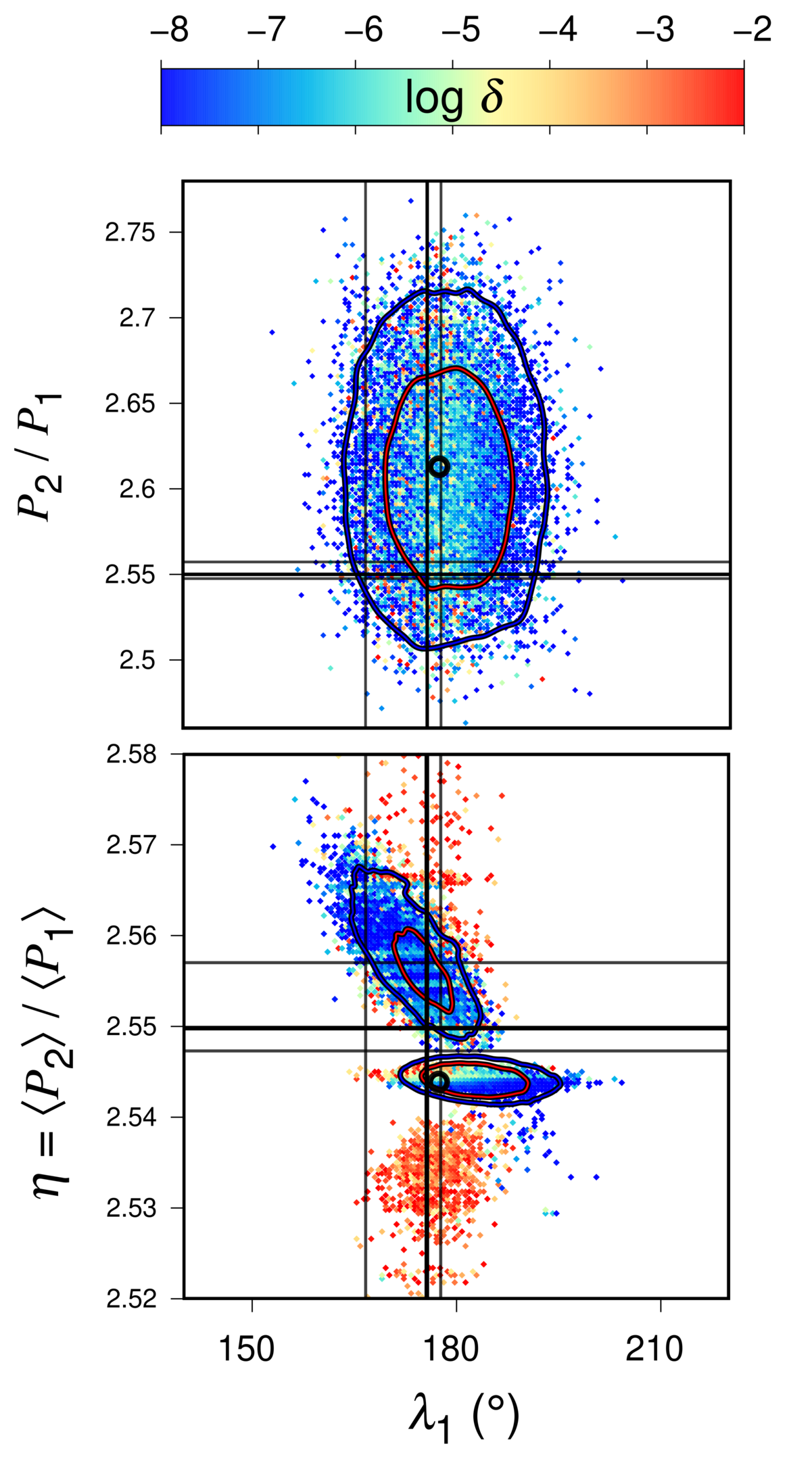}
  \caption{Confidence levels for ($\lambda_1, P_2/P_1$) (\textit{top}) and ($\lambda_1, \eta$) (\textit{bottom}). Blue and red curves represent the $1\sigma$ and $2\sigma$ regions, respectively. The straight lines indicates the $P_1$ and $\lambda_1$ values corresponding to the Keplerian fit and its $\pm1\sigma$ errors (Table~\ref{Table 1}), while the circle marks the Newtonian best-fit value (Table~\ref{Table 2}). The colour scale corresponds to the diffusion index on a logarithmic scale (Eq. \ref{diffusionindex}).}
  \label{confidencelevels}
\end{figure}

The confidence levels in the corner stability maps are derived from the stable initial conditions of the posterior sub-samples of the Newtonian fit (Table~\ref{Table 2}). Figure~\ref{confidencelevels} shows all points of the posterior distribution coloured according to its stability index, $\DeltaIn$ (Eq.~\ref{diffusionindex}). We determined the confidence levels using the non-parametric kernel density estimation (KDE) method, which generates a kernel (e.g. a Gaussian distribution) for every stable initial condition, i.e. those with $\DeltaIn < 10^{-5}$, allowing for a better estimation of the probability density function and, consequently, smoother curves \citep{parzen1962estimation,chen2017tutorial}. We represent the $1 \sigma$ and $2 \sigma$ of the stable posterior sub-samples, which surround about $68\%$ and $95\%$ of the stable initial conditions, respectively.

Figure \ref{confidencelevels} also provides a comparison between confidence levels considering the ratio of osculating orbital periods, $P_2/P_1$ (top panel), and the mean orbital period ratio, $\eta =\Pms/\Pm$ (bottom panel). As in Fig.~\ref{ratiocomparison}, the mean orbital periods significantly modify the posterior distribution and the corresponding confidence levels. The curves in the top panel exhibit a two-dimensional Gaussian behaviour. In the bottom panel, the KDE confidence levels surround two distinct regions that represent the most probable configurations for the system. The Keplerian fit is used here only as a compact representation of the dominant RV periodicity and phase (Fig.~\ref{fig:vrphase}). The dynamical interpretation is instead based on the Newtonian posterior samples and on the proper frequencies measured from the corresponding integrations.

\subsection{Results}
\label{coplanresults}

Figure~\ref{CornerAll} shows the corner stability maps for the parameters ($\eta, \lambda_1, e_1, \omega_1$).
Horizontal and vertical lines indicate the Keplerian best fit values, $\eta \sim 2.552$ and $ \lambda_1 = 175.69^\circ$, respectively (Table~\ref{Table 1}).
As expected, the best-fit solutions fall within the red level curves in all stability maps.
We can clearly identify regions with low diffusion and thus determine the sets of orbital parameters that guarantee long-term orbital stability for the planet and simultaneously best fit the observational data.

We observe that for all combinations of the orbital parameters, blue stable regions exist within the confidence levels.
However, while for the subset ($\omega_1, e_1$) there is a wide range of values that match the observations, the confidence levels are very constrained and sharp for the subsets involving $\eta$ or $\lambda_1$.
Moreover, these levels extend along straight lines near the values obtained with the Keplerian best fit (Table~\ref{Table 1}).
We hence confirm that the mean orbital periods and the mean longitudes are well constrained and that their values are close to the Keplerian ones.

\begin{figure*}
\centering
  \includegraphics[width=\textwidth]{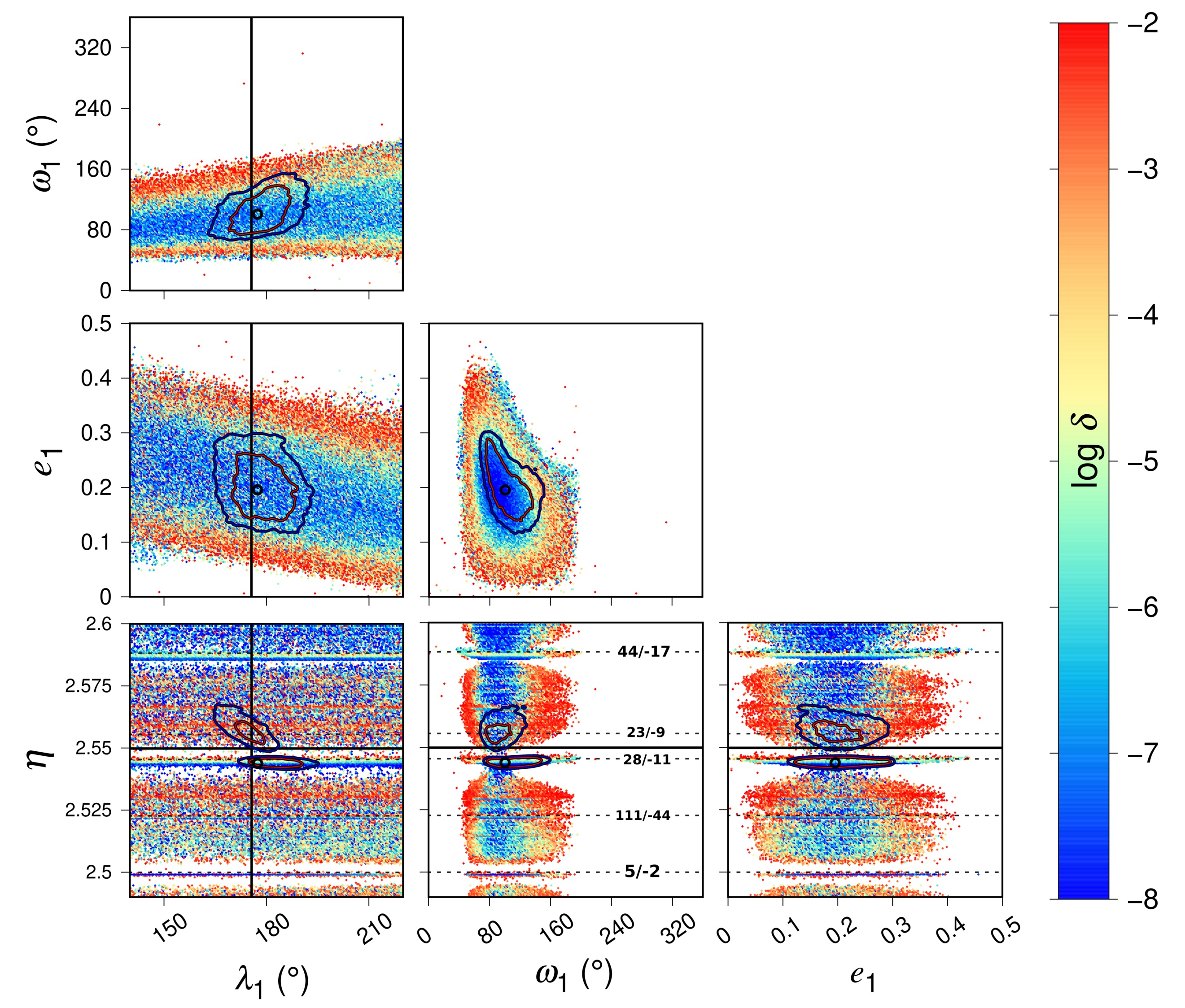}
  \caption{Corner stability maps for all surviving initial conditions. The solid lines and circles correspond to the Keplerian fit and the Newtonian best stable fit, respectively. The confidence levels represent the $1\sigma$ (red) and $2\sigma$ (blue) regions of the posteriors. The nominal locations of commensurabilities with the binary, computed from $\eta$, are indicated by dashed lines; these lines do not represent analytic resonance-width estimates. The colour scale corresponds to the diffusion index in logarithmic scale (Eq.~\ref{diffusionindex}).}
  \label{CornerAllzoom}
\end{figure*}

To obtain a more detailed view of the stability regions within the confidence levels, in Figure \ref{CornerAllzoom}, we performed a zoomed Monte Carlo scan with $6 \times 10^6$ initial conditions, restricting the ratio of the mean orbital periods to $\eta \in [2.49,2.6]$ and the mean longitude to $\lambda_1 \in [140^\circ,220^\circ]$, thereby zooming in on the corner stability maps around the confidence levels derived from the posterior sub-samples \citep{cheng2025}. We therefore explored the phase space in more detail around the regions consistent with the observational and stability constraints.
In the vicinity of the best fit ($\lnL = -1230.3$), we find a large number of stable initial conditions within the $1\sigma$ and $2\sigma$ confidence levels.

Figure~\ref{CornerAllzoom} shows that the eccentricity is constrained within $0.1 \lesssim e \lesssim 0.35$ and the argument of pericentre within $70^\circ \lesssim \omega_1 \lesssim 140^\circ$.
In particular, at the centre of the ($\omega_1, e_1$) map, we observe that these values belong to a large stability island in the middle of a chaotic region.
The trajectories within this area mostly correspond to orbital solutions that show libration of the angle $\Delta \varpi = \varpi_1 - \varpi_2$ around $0^\circ$, where the longitudes of the pericentres are given by $\varpi_1 = \Omega_1 - \omega_1$ and $\varpi_2 = \Omega_2 + \omega_2$.
For these initial conditions, the orbits of the planet and $\nu$-Oct~B have aligned pericentres; that is, they present a secular equilibrium which prevents close encounters between the two bodies. 

The picture is much less clear for the ($\lambda_1, \eta$) map, since stable regions alternate with unstable regions. Although there are initial conditions with low diffusion indices in the best-fit region, the region globally exhibits chaotic behaviour. However, we can identify horizontal straight lines of stability in the middle of highly unstable white regions, which correspond to MMRs.
In this map, the confidence levels are divided into two regions: an upper region with $\eta \sim 2.555$ and a lower region with $\eta \sim 2.544$, where the latter clearly has smaller diffusion indices. The 28/$-$11~ MMR lies within this region, and its separatrix defines the gap between the two stable regions.

The impact of MMRs is also striking in the ($\omega_1, \eta$) and ($e_1,\eta$) maps. The stability increases along the resonant orbital periods (in particular for the 28/$-$11~MMR), while around these resonant islands there is a significant reduction in stability (white and red regions) due to the presence of a separatrix between the resonant and circulation regimes. Because the perturber is massive and eccentric, the local resonant web is strongly overlapped. We therefore did not attempt to assign analytic widths to the resonances in these projections. Instead, the dashed lines mark nominal commensurabilities in proper-frequency space, and the effective resonant domains are identified numerically from the low-diffusion structures and from the frequency decomposition discussed in Appendix~\ref{sec:fundfreq}.

\begin{figure}
        \centering
        \includegraphics[width=\linewidth]{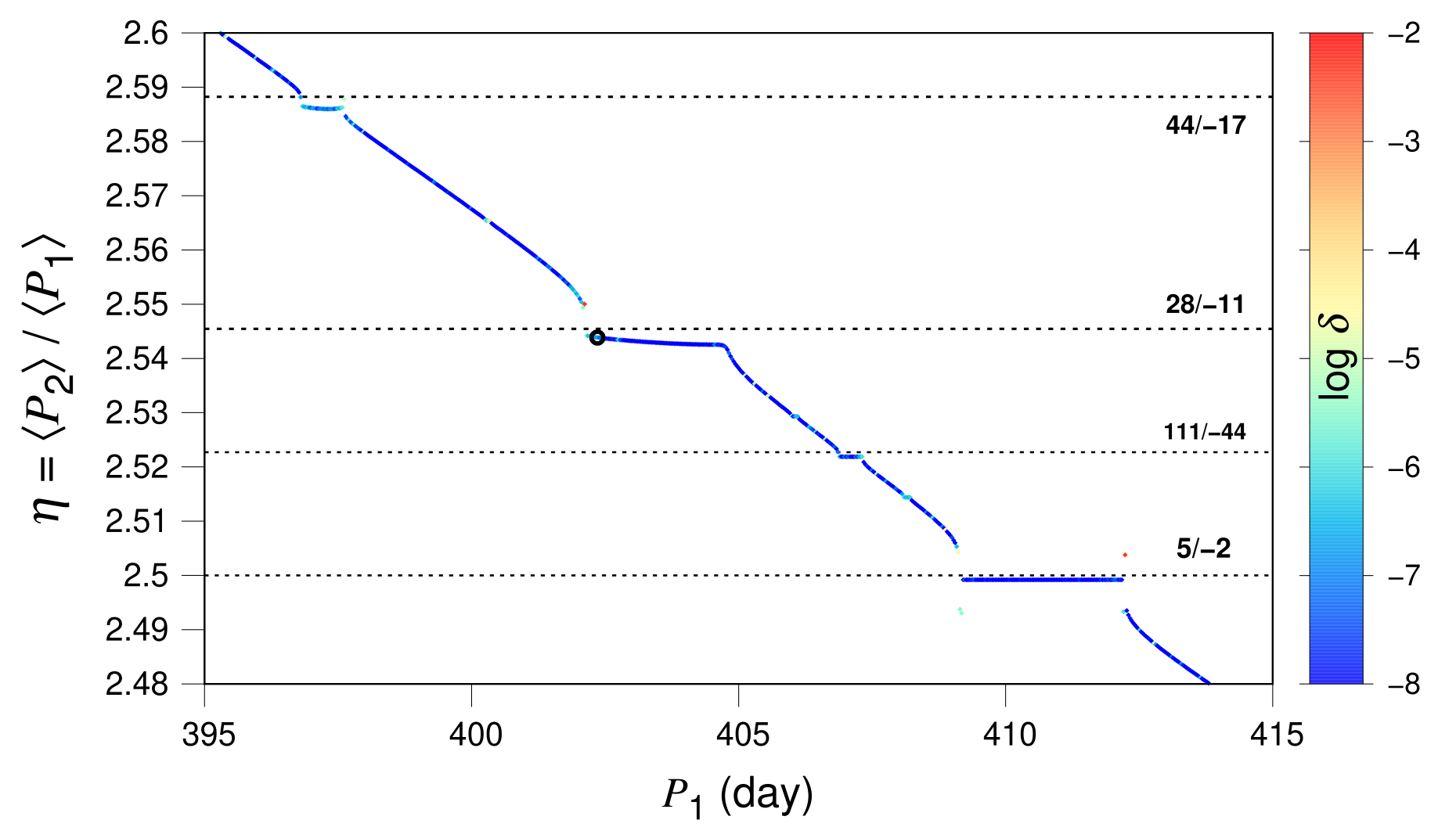}
        \caption{Frequency analysis of the mean longitudes as a function of the osculating initial orbital period, $P_1$, while fixing $\lambda_1 = 177.44^\circ$ and the remaining orbital parameters to the Newtonian best stable-fit initial condition (Table~\ref{Table 2}), indicated by the black circle. The horizontal dashed lines correspond to the location of the nominal 28/$-$11, 44/$-$17, 5/$-$2, and 111/$-$44~MMRs.}
\label{fig:ress3112}
\end{figure}

To confirm the resonant nature, we have considered the Newtonian best stable fit solution (Table~\ref{Table 2}) and performed a frequency analysis by varying the orbital period, $P_1$, while fixing all the remaining orbital parameters.
Figure~\ref{fig:ress3112} shows the ratio of the mean orbital periods, $\eta$, as a function of the initial orbital period of the planet, $P_1$.
The frequency analysis shows elliptic stability points at given frequencies (horizontal thresholds), which reveal the dynamical signatures of the corresponding commensurabilities \citep[see][]{laskar1993frequency}, namely the 5/$-$2, 28/$-$11, 44/$-$17, and 111/$-$44~MMRs. The apparent horizontal extent of each plateau in this one-dimensional scan is a projected stability interval along the chosen osculating $P_1$ coordinate, not the physical width of the resonance in action space.

\begin{figure*}
        \centering
        \includegraphics[width=\textwidth]{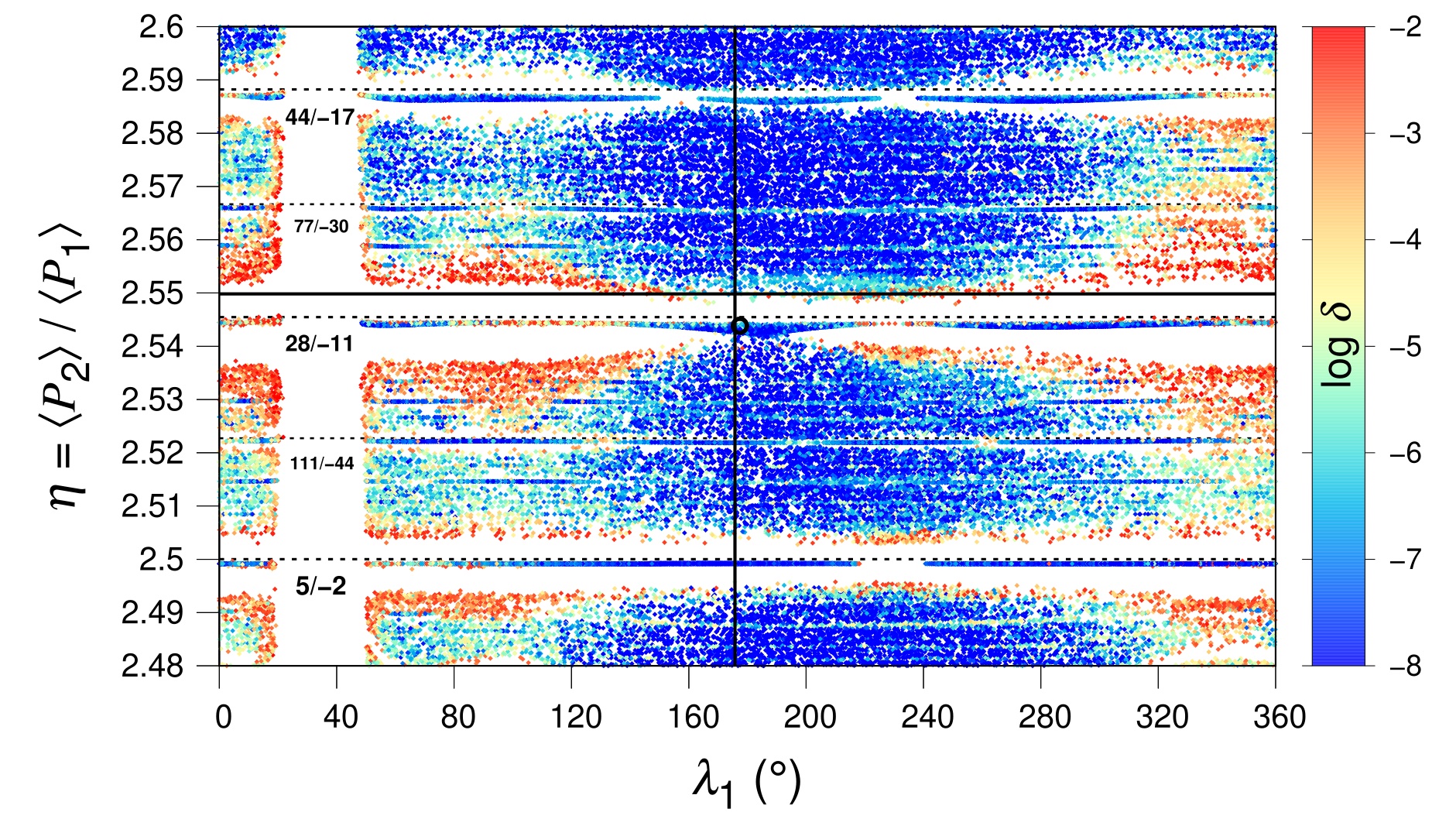}
        \caption{Stability map for ($\lambda_1, \eta$) obtained by varying the initial values of ($\lambda_1, P_1$), while fixing the remaining orbital parameters of the Newtonian best stable-fit initial condition (Table~\ref{Table 2}), indicated by the black circle. The solid and dashed lines represent the Keplerian Fit and the nominal locations of the MMRs, respectively.}
\label{fig:ressplanarv2}
\end{figure*}

For a clearer view of the stability within the resonant regions, we performed another frequency analysis by varying the initial mean longitude, $\lambda_1$, and the initial orbital period, $P_1$, while fixing all the remaining orbital parameters.
Figure ~\ref{fig:ressplanarv2} shows the corresponding ($\lambda_1, \eta$) stability map.
In this case, each pixel corresponds to a single pair of initial conditions ($\lambda_1, P_1$) and we can clearly observe the lines corresponding to the MMRs.  

A detailed analysis of Figs.~\ref{CornerAllzoom},~\ref{fig:ress3112}, and \ref{fig:ressplanarv2} yields the following conclusions: 

\begin{enumerate}
\item We do not find posterior-supported stable solutions centred on the low-order 5/$-$2~MMR, although the osculating orbital period is relatively close to this ratio. When we compute the mean orbital period ratio, $\eta$, it becomes clear that this MMR lies beyond the confidence levels of the posteriors.
\item The frequency maps reveal the dynamical signatures of the higher-order 28/$-$11 and 44/$-$17 commensurabilities in phase space. Higher-order resonances also exist in small regions of phase space, for example, the 77/$-$30 and 111/$-$44~MMRs. 

\item In the region of the best fit, there are also many non-resonant stable trajectories. The mean orbital solution obtained with the Keplerian fit (Table~\ref{Table 1}) lies in one of these regions between the 28/$-$11 and the 23/$-$9~MMRs. 
\end{enumerate}

\section{Inclined case}
\label{sec4}

\begin{figure*}
\centering
  \includegraphics[width=\textwidth]{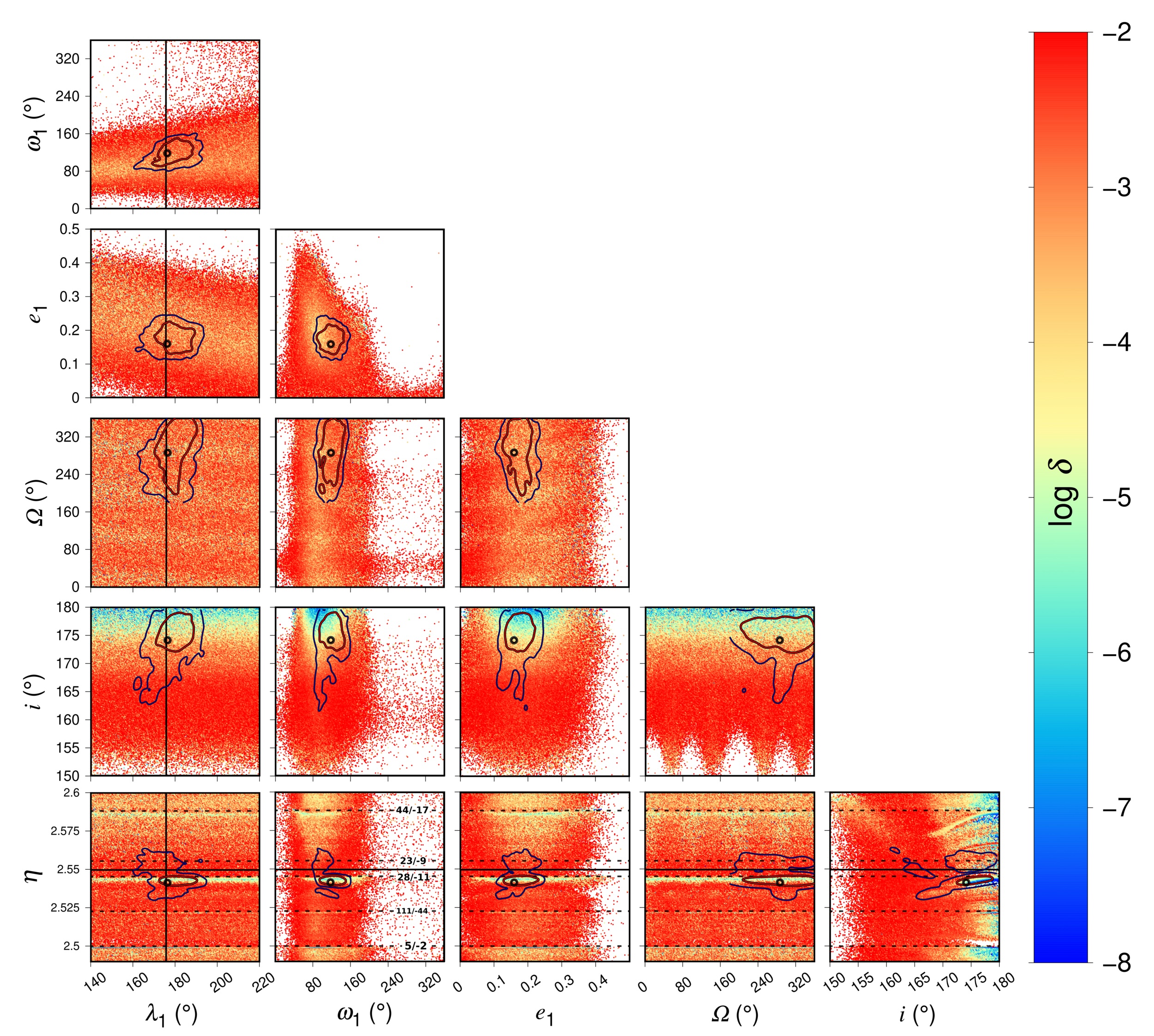}
  \caption{Corner stability maps for the inclined case, showing all surviving initial conditions. The solid lines represent the Keplerian fit (Table~\ref{Table 1}), while the circles mark the maximum $\lnL$ value of all stable solutions from the inclined posterior sub-sample. The confidence levels represent the $1\sigma$ (red) and $2\sigma$ (blue) regions. The nominal locations of commensurabilities with the binary, computed from $\eta$, are represented by dashed lines.}
  \label{fig7:4Corner}
\end{figure*}

\begin{figure*}
\centering
  \includegraphics[width=\textwidth]{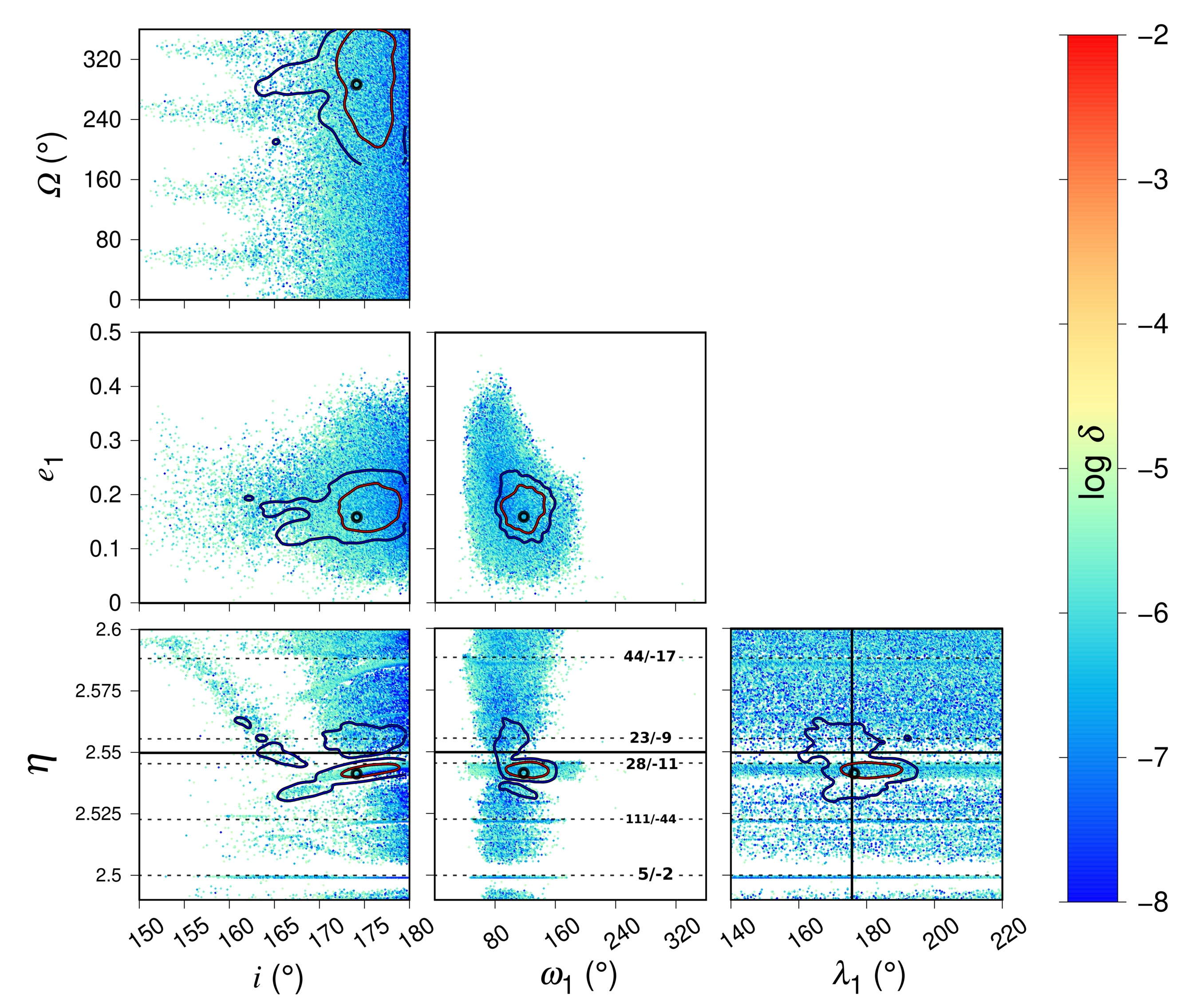}
  \caption{Corner stability maps for the inclined case considering only initial conditions with $\log \DeltaIn < -5$. The solid lines represent the Keplerian fit (Table~\ref{Table 1}), while the circles mark the maximum $\lnL$ value of all stable solutions from the inclined posterior sub-sample. The confidence levels correspond to the $1\sigma$ (red) and $2\sigma$ (blue) regions. The nominal locations of commensurabilities with the binary are represented by dashed lines.}
  \label{fig:2Cornerdiff}
\end{figure*}

\citet{gozdziewski2013} studied the stability of the $\nu$-Oct system for several inclination values. They showed that instability dominates phase space, but stable orbits can be found for mutual inclinations greater than about $165^\circ$. \citet{cheng2025} also explored the inclined case in more detail and determined a similar stability boundary for the inclination of the planet's orbit ($i > 160^\circ$). Therefore, in this section, we relax the previous constraint of a retrograde coplanar orbit and explore the phase space by considering inclined orbits for the planet.
We still assume that the orbital parameters of $\nu$-Oct~B are well determined and adopt the reference values provided in Table~\ref{Table 2}.
As a result, in addition to the four orbital parameters ($P_1, \lambda_1, e_1, \omega_1$), we explored the inclination to the line of sight, $I_1$, and the longitude of the node, $\Omega_1$.

As we changed the values of $I_1$, we also expected the mass of the planet to change, since $K_1 \propto m_1 \sin I_1$ \citep[e.g.][]{murraycorreia2010}.
Although the mass is not a critical parameter (unless $\sin I_1 \ll 1$), to maintain some continuity with the coplanar case studied previously, we adopted $m_1 = 2.058 $~M$_\mathrm{Jup} / \sin I_1 $, so that for $I_1 = 108.2^\circ$, we retrieved the mass of the best stable coplanar fit (Table \ref{Table 2}).

To better understand the dynamics, we adopted the orbit of $\nu$-Oct~B as the reference plane, instead of using the plane of the sky.
We then directly scanned the mutual inclination, $i$, between the two orbital planes, together with the relative longitude of the node, $\Omega$, measured along the orbit of $\nu$-Oct~B.
Following \citet{giuppone2012dynamical}, we provide the conversion between the two reference frames in Appendix~\ref{framesection} for completeness.

We generated $25 \times 10^6$ different initial conditions for the six unknown parameters using random uniform distributions spanning the intervals: $P_1 \in [330,451]$~day, $\lambda_1 \in [140^\circ,220^\circ]$, $e_1 \in [0,0.5]$, $\omega_1 \in [0^\circ,360^\circ]$, $i \in [150^\circ,180^\circ]$, and $\Omega \in [0^\circ,360^\circ]$.
We restricted the orbital period and the mean longitude following the information from the coplanar case (Sect.~\ref{coplanresults}) and the mutual inclination according to the results from \citet{gozdziewski2013} and \citet{cheng2025}. 

Figure~\ref{fig7:4Corner} shows the corner stability map (Sect.~\ref{CSMap}) with confidence levels (Sect.~\ref{sec:confidence}) for the six free parameters.
We determined the confidence levels from the inclined posterior sub-samples \citep{cheng2025}, and the black circles represent the best fit that has a low diffusion index, corresponding to $\lnL = -1228.71$ with $i = 174.1^\circ$ and $\Omega = 286.7^\circ$. The remaining orbital parameters, $P_1 =  409.7$~day, $\lambda_1 = 176.3^\circ$, $e_1=0.16$, and $\omega_1 = 118^\circ$, are relatively close to those of the Newtonian best stable fit for the retrograde coplanar model (Table~\ref{Table 2}).

As in \citet{gozdziewski2013}, we observe that unstable orbits dominate the phase space, but we also confirm that some stability regions exist for mutual inclinations larger than $160^\circ$. In particular, from the $\eta$ dependent maps, we observe that the vicinity of the 28/$-$11~MMR corresponds to the region with the lowest diffusion in phase space.

Since we explore all the free parameters in Fig.~\ref{fig7:4Corner}, we can determine how the inclination correlates with them. The confidence levels are restricted to well-defined regions. However, as there are many unstable solutions that somehow overshadow the few stable regions, in Fig.~\ref{fig:2Cornerdiff}, we filter our results by showing only the solutions with a stability index $\DeltaIn < 10^{-5}$. 
Moreover, we restricted the corner stability map to the most important combinations of orbital parameters.
 
In Fig.~\ref{fig:2Cornerdiff}, one striking result is that we clearly observe stable initial conditions populating regions that were previously classified as mostly unstable.
This means that, although stability is unlikely in those regions, some combinations of the orbital parameters survive there.
Filtering our results using even smaller diffusion indices ($\DeltaIn < 10^{-7}$) further shows that most of the initial conditions are near the 28/$-$11~MMR and in the secular equilibrium configuration.

In Fig.~\ref{fig:2Cornerdiff}, we also observe that the more heavily populated stability regions cluster around the same combination of parameters that we obtained for the retrograde coplanar case (Sect.~\ref{coplanresults}): a large number of trajectories have arguments of pericentre within $60^\circ \lesssim \omega_1 \lesssim 150^\circ$ and align with the resonant ratios, mainly for the 28/$-$11~MMR. This feature is also clearly visible in the ($i, \eta$) map, which shows that MMR regions are resilient even for inclined orbits. In this map, the confidence levels are well-constrained and centred around the nominal 28/$-$11~MMR. Thus, although the confidence levels are divided into two distinct regions, as in the coplanar case (Sect.~\ref{coplanresults}), for the inclined case, the $1\sigma$ level surrounds only the 28/$-$11 MMR region, showing that this area exhibits the highest likelihood and stability in phase space.

Finally, in the ($i, e_1$) map, we observe that, as the inclination decreases, there is a clear preference for eccentricities clustering around 0.2, which contrasts with the wider interval $0.1 \lesssim e_1 \lesssim 0.35$ in the retrograde coplanar case (Sect.~\ref{coplanresults}). 
Interestingly, $e_1 \sim 0.2$ matches the centre of the stability island in the ($\omega_1, e_1$) map, which corresponds to trajectories that show libration in the angle $\Delta \varpi$ around $0^\circ$ (aligned orbits).

\section{Conclusion}
\label{sec5}

We investigated the possible stable configurations of the planet in the $\nu$~Octantis system that comply with the observational constraints.
We first analysed the available RV data and determined the parameters that are well, moderately, or loosely constrained.
We realised that, although most orbital parameters of the planet undergo large oscillations in time, their average values are very well defined and can thus be used to restrict the phase space of possible orbital configurations.

Because there are many unconstrained orbital parameters, we introduced a novel method, which we refer to as corner stability maps, that explores the stability of the system using a Monte Carlo approach.
We then projected all combinations of parameters onto 2D maps, which allowed us to easily determine the correlations between them. We drew confidence levels based on the maximum likelihood, $\lnL$, and the long-term stability of the orbital solutions obtained, which relates the reliability of each orbital parameter determination.

We inferred the stability of the orbital solutions using frequency analysis.
This technique not only quantifies the diffusion of the orbits, but also allows us to extract the mean values of the orbital parameters, which can be compared with the observational constraints.
In addition, this analysis clearly identifies the frequencies corresponding to MMRs, providing further insight into the dynamics of the system.

We explored two cases in detail: the retrograde coplanar case and the inclined case. In both cases, we find a wide variety of possible stable configurations. One possibility is that the planet is in a secular equilibrium with its orbit aligned and precessing together with that of $\nu$-Oct~B (libration of $\Delta \varpi$ around $0^\circ$). 

%At first glance, this pericentre alignment did not occur..
At first glance, libration of $\Delta \varpi$ did not occur for more than half of the stable initial conditions. However, we find that this apparent non-librating regime is caused by conjunctions between the planet and $\nu$-Oct~B, resulting in repeated quasi-periodic close encounters between the two bodies. During these encounters, the osculating $\varpi_1$ of the planet exhibits large jumps, producing an instantaneous artificial circulation of $\Delta \varpi$. We predominantly observed this behaviour near or within MMR regions. After filtering out the short-period frequencies above $n_2/5$, the secular trend becomes clearly visible, and we find that about $97\%$ of the stable initial conditions in the posterior distributions exhibit libration of $\Delta \varpi$. An animation representing this filtering is available online.

We do not find posterior-supported stable solutions centred on the 5/$-$2~MMR proposed by \citet{gozdziewski2013}. Instead, the stable posterior sub-samples are concentrated near the higher-order 28/$-$11, 23/$-$9, 77/$-$30, 87/$-$34, 71/$-$28, and 89/$-$35 commensurabilities, which are consistent with the observations in the sense of proper-frequency space. Although weak, high-order resonances have been previously reported for some Trans-Neptunian objects \citep[e.g.][]{crompvoets2022ossos}.

For the inclined case, we show that non-coplanar orbits consistent with the observations exist for mutual inclinations above $160^\circ$. However, about 90$\%$ of the possible low-diffusion orbital solutions have $i \ge 170^\circ$, which is consistent with the near coplanar case. These initial conditions are mostly in the region of influence of the 28/$-$11 MMR. We also identified some orbits trapped in the 71/$-$28 and 89/$-$35~MMRs, which were not found in the coplanar case.

Table~\ref{Tablestats} presents the statistical distribution of all stable initial conditions for the posterior sub-samples ($\DeltaIn < 10^{-5}$) in the two regions of the ($\lambda_1,\eta$) stability map (Figs.~\ref{CornerAllzoom} and \ref{fig:2Cornerdiff}). The percentages of initial conditions in the regions of influence of the most probable MMRs are shown, while the percentage corresponding to orbits stabilised exclusively by libration of $\Delta \varpi$ is shown separately. The remaining small percentage of high-order MMRs and unknown configurations is classified as `Other'.

The region of influence of the 28/$-$11~MMR is the most stable in the entire phase space and contains the Newtonian best stable fits for both the coplanar and inclined cases. Both fits exhibit libration of $\Delta \varpi$. Therefore, the planet is most likely on a nearly coplanar orbit in the region of influence of the 28/$-$11 commensurability, with aligned pericentres.

Further observations, such as those by the {\it Gaia} satellite \citep{gaia2016}, may further constrain the possible stable orbits in the system.
Our methodology for searching for orbital stability in compliance with the observational constraints is very general and can therefore be applicable to other planetary systems, in particular those subject to strong gravitational interactions and exhibiting large variations in orbital parameters.

\begin{table*}[h!]
\centering
\caption{Statistical distribution of the stable initial conditions for the posterior sub-samples ($\DeltaIn < 10^{-5}$).}
\label{Tablestats}
\renewcommand{\arraystretch}{1.3}
\begin{tabular}{c|cccccc|cccc}
\toprule
\toprule
\multicolumn{1}{c|}{{\bf Posterior sub-samples}} &  & \multicolumn{4}{c}{{\bf Lower Region ($\eta \sim 2.544$)}} &  & \multicolumn{3}{c}{{\bf Upper Region ($\eta \sim 2.555$)}} \\ 
\midrule
\multirow{3}{*}{\begin{tabular}[c]{@{}c@{}}Coplanar stable \\ solutions: 14984\end{tabular}} 
&  & \multicolumn{4}{c}{7972 (53.17\%)} &  & \multicolumn{3}{c}{7012 (46.83\%)} \\
&  & Only $\Delta\varpi = 0$ & 28/$-$11 & 89/-35 & Other &   & Only $\Delta\varpi = 0$ & 23/$-$9 & Other \\
&  & 9.35\% & 89.34\% & 0.1\% & 1.21\% &    & 76.25\% & 14.33\% & 9.42\% \\ 
\midrule
\multirow{3}{*}{\begin{tabular}[c]{@{}c@{}}Inclined stable \\ solutions: 1179\end{tabular}} 
&  & \multicolumn{4}{c}{963 (81.68\%)} &  & \multicolumn{3}{c}{216 (18.32\%)} \\
&  & Only $\Delta\varpi = 0$ & 28/$-$11 & 89/-35 & Other &  & Only $\Delta\varpi = 0$ & 23/$-$9 & Other \\
&  & 29.8\% & 48.91\% & 11.73\% & 9.56\% &  &  58.79\% & 28.24\% & 12.97\% \\ 
\bottomrule
\end{tabular}
\tablefoot{The solutions were divided in two regions of the $(\lambda_1,\eta)$ map. Besides the classification of the stable mechanisms, the corresponding number of stable solutions in each region is also shown.}
\end{table*}

\vspace{-0.2cm}
\section*{Data availability}
To assess the pericentre alignment, we filtered out the short-period frequencies of an initial condition close to the Newtonian best fit and represented it in a video considering a heliocentric reference frame co-precessing with $\nu$ Oct B. The animation is available online via \url{https://repositorio.unesp.br/handle/11449/328080}.

\begin{acknowledgements}
We are grateful to V. M. Oliveira and G. A. Caritá for enlightening conversations on the topics of this article.
This work was supported by the grants FAPESP/2021/11982-5, FAPESP/2022/08716-4, FAPESP/2023/02528-4 and FAPESP/2024/10557-7 of S\~ao Paulo Research Foundation, National Council for Scientific and Technological Development (CNPq), Brazil, Grant 305997/2025-4, and by the FCT - Funda\c{c}\~ao para a Ci\^encia e a Tecnologia, I.P., Portugal, through the CFisUC project UID/04564/2025 (with DOI identifier 10.54499/UID/04564/2025) and national funds, and by the European Regional Development Fund (ERDF) through the Innovation and Digital Transition Thematic Program (COMPETE 2030), of Portugal 2030, and by the European Union - Operation No. 14981 - COMPETE2030-FEDER-00860300 SPACE. M.H.L.\ and H.W.C.\ were supported in part by the Hong Kong RGC grant 17309323. The computational resources were supplied by the Center for Scientific Computing (NCC/GridUNESP) of the São Paulo State University (UNESP). 
\end{acknowledgements}

\vspace{-0.7cm}
\bibliographystyle{aa} % style aa.bst
\bibliography{bib}

\appendix

\section{$\nu$-Oct~B reference frame}
\label{framesection}

Radial velocity measurements use the plane of the sky as reference frame (the $z$-axis being aligned with the observer's vantage point).
However, for a better understanding of the dynamics of the system, it is preferable to use the orbital plane of the binary stars as a reference, which directly gives us the mutual inclination, $i$.
The transformation between the two reference frames can be obtained using spherical trigonometry relations \citep[e.g.][Fig.~1]{giuppone2012dynamical}, which make use of two additional angles: the difference of node longitudes in the plane of the sky $\Delta \Omega = \Omega_1 - \Omega_2$ and the difference of the argument of the pericentre of the planet in both reference frames $\Delta \omega = \omega_1 - \omega$. 

To study the system with respect to the orbital plane of binary stars ($\Omega, i$), it is necessary to use the following relations
\begin{align}
& \cos{i} = \cos{I_2} \cos{I_1} + \sin{I_2} \sin{I_1} \, \cos{\Delta \Omega} \ , \label{eq1} \\
& \cos{\Delta \omega} = \frac{(\cos{I_2} - \cos{I_1} \, \cos{i})}{(\sin{I_1} \, \sin{i})} \ , \label{eq5} \\
& \cos{\Omega} = \cos{\Delta \Omega} \,\cos{\Delta \omega} - \sin{\Delta \Omega} \, \sin{\Delta \omega} \, \cos{I_1} \ . \label{eq6}
\end{align}
We further need to note that when $0^\circ < \Omega_1 < 180^\circ$, it implies that  $0^\circ < \Delta \omega < 180^\circ$ and $0^\circ < \Delta \Omega < 180^\circ$, and for $180^\circ < \Omega_1 < 360^\circ$, we have $180^\circ < \Delta \omega < 360^\circ$ and $180^\circ < \Delta \Omega < 360^\circ$. 

If we wish to convert the relative parameters back to the plane of the sky ($\Omega_1, I_1$), the transformations are
\begin{align}
& \cos{I_1} = \cos{I_2} \cos{i} - \sin{I_2} \sin{i} \, \cos{\Omega} \ , \label{eq4}\\
& \cos{\Delta \omega} = \frac{(\cos{I_2} - \cos{I_1} \, \cos{i})}{(\sin{I_1} \, \sin{i})} \ , \label{eq2}\\
& \cos{\Delta \Omega} = \cos{\Omega} \,\cos{\Delta \omega} + \sin{\Omega} \, \sin{\Delta \omega} \, \cos{i} \ . \label{eq3}
\end{align}

\section{Fundamental Frequencies}
\label{sec:fundfreq}

We performed a detailed study of the quasi-periodic approximation, obtained for the Newtonian best stable fit (Table \ref{Table 2}), by frequency analysis for the time interval $\left[0,120\right]$~kyr. This longer interval is used only for representative solutions in order to resolve the slow secular terms; it is not the computationally expensive scan used for all Monte Carlo initial conditions.
The first 50 terms in descending order are given for the planet and $\nu$-Oct~B, respectively, in Tables \ref{tab:freq_planet} and \ref{tab:freq_binary} for the variables $\xi=a\, \mathrm{exp}(i\lambda)$ and $z=e \, \mathrm{exp}(i\varpi)$.
The frequencies of the dominant terms for $\xi$ and $z$ correspond, respectively, to the mean motion and the pericentre precession frequency.
In Table \ref{tab:freq_ref}, we then obtain the mean motion of the planet, $n_1$, of $\nu$-Oct~B, $n_2$, and the pericentre precession frequency $g$.
We observe that the precession frequencies are identical, given the accuracy of frequency analysis for the planet and $\nu$-Oct~B, corresponding to the libration of $\Delta \varpi$ previously mentioned.
However, we cannot find a Diophantine equation of the kind $jn_1+kn_2+lg=0$ with $j$, $k$, $l$ integers, which could indicate a mean motion resonance.
On the contrary, it is possible to obtain such a relation ($11n_1+28n_2-39g=0$) for the frequency analysis of nearby solutions trapped in the 28/$-$11 MMR. This is the criterion used in the manuscript to identify the 28/$-$11 region: it relies on the proper frequencies, not on the instantaneous osculating period ratio.
It is possible to identify other terms as a linear combination with integer coefficients of the three frequencies $n_1$, $n_2$, and $g$, like in \cite{Laskar1990,Laskar2004}, but not all.
For instance, the frequency of the second term for the decomposition $\xi$ of the planet is close to $n_1$ and can be written as $n_1-\sigma$ with sigma the deviation to $n_1$, and the ninth term for the decomposition $z$ of the planet can be designed by a supplementary frequency $f$.
With the add of these two supplementary frequencies $\sigma$, and $f$ as done in \cite{Laskar1990}, we can identify all the terms as a linear combination $jn_1+kn_2+lg+mf+p\sigma$ with $j$, $k$, $l$, $m$, $p$, integers verifying the d'Alembert relations $j+k+l+m=0$ \citep[e.g.][]{Laskar1995}.
These five ``fundamental'' frequencies $n_1$, $n_2$, $g$, $f$, $\sigma$ are indicated in Table \ref{tab:freq_ref}.
Adopting a similar approach to that used by \citet{Laskar1990} for a secular resonance, we investigate the 28/$-$11 MMR initial conditions and determine that $\sigma$ corresponds to the libration frequency. In this strongly perturbed regime, we therefore use the numerical frequency relation and the associated libration frequency as the practical resonance diagnostic. 
We can also note that $\sigma$ corresponds to a difference of frequencies and its associated integer coefficient $p$ is then not involved in the d'Alembert relations.

\begin{table}
\caption{\label{tab:freq_ref} Fundamental frequencies identified in the frequency decompositions of the Newtonian best stable fit solution on $\left[0:120\right]\mathrm{kyr}$.}

\begin{center}
\begin{tabular}{c c c} \hline \hline
\noalign{\smallskip}
&{\bf Frequency}  & {\bf Period}   \\ 
&{\bf (deg/day)}  & {\bf (day)}  \\ \hline 
\noalign{\smallskip}
${n_1}$ & -0.87225967830519513    & 412.721129904  \\ 
${n_2}$ & 0.34288883489243022   & 1049.90295211    \\  
${g}$ & 0.000077981832018774973    & 4616459.89432 \\ 
${f}$ & -1.3845645296718281    & 260.009549779  \\ 
${\sigma}$ & 0.00287790080617667    & 125091.177301 \\ 
\hline
\end{tabular}
\end{center}
\end{table}

\newcolumntype{L}[1]{>{\raggedright}p{#1}}
\newcolumntype{R}[1]{>{\raggedleft}p{#1}}
\newcolumntype{C}[1]{>{\centering}p{#1}}

\begin{table*}
\caption{\label{tab:freq_binary}First 50 terms of the frequency decomposition $\sum_{k=1}^{50}A_ke^{i\left(\nu_k t+\phi_k\right)}$ of $\xi$ (a) and $z$ (b) for $\nu$-Oct~B on $\left[0:120\right]\mathrm{kyr}$ for the solution obtained for the best Newtonian fit for the coplanar case (Table~\ref{Table 2}).}
\centering
\subfloat[$\xi=a \mathrm{exp}(i\lambda)$]{
\begin{tabular}{@{}C{3cm}R{2.5cm}R{1.3cm}r@{}}
\hline
 & $\nu_k$ (deg/day) & $10^6\times A_k$ & $\phi_k$ (deg) \\
\hline
$n_1$ & -0.8722596783    &               913383          & -88.70 \\ 
$n_1$-$\sigma$ & -0.8751375791   &               562831          & -86.14 \\ 
$n_1$+$\sigma$ & -0.8693817775   &               541711          & 88.74 \\ 
$n_1$+2$\sigma$ & -0.8665038767          &               157720          & -93.83 \\ 
$n_1$-2$\sigma$ & -0.8780154799          &               121463          & -83.57 \\ 
$n_1$+3$\sigma$ & -0.8636259757          &                51495          & 83.61 \\ 
-$n_1$+2$n_2$ & 1.5580373482     &                49384          & -162.88 \\ 
$n_1$-3$\sigma$ & -0.8808933807          &                31819          & -81.01 \\ 
-$n_1$+2$n_2$+$\sigma$ & 1.5609152487    &                30066          & -165.43 \\ 
-$n_1$+2$n_2$-$\sigma$ & 1.5551594473    &                29181          & 19.69 \\ 
$n_2$ & 0.3428888350     &                28289          & 54.21 \\ 
$n_1$-$n_2$+$g$ & -1.2150705314          &                27926          & 112.06 \\ 
$n_1$+$n_2$-$g$ & -0.5294488251          &                27805          & -109.46 \\ 
2$n_2$-$g$ & 0.6856996877        &                21381          & -146.54 \\ 
$f$ & -1.3845645295      &                20208          & 176.36 \\ 
$n_1$+$n_2$-$g$-$\sigma$ & -0.5323267261         &                17224          & -106.89 \\ 
-$n_1$+$g$+$f$+$\sigma$ & -0.5093489692          &                17042          & -22.52 \\ 
$n_1$-$n_2$+$g$-$\sigma$ & -1.2179484319         &                16885          & 114.62 \\ 
$n_1$+$n_2$-$g$+$\sigma$ & -0.5265709240         &                16672          & 67.96 \\ 
$n_1$+4$\sigma$ & -0.8607480738          &                16647          & -98.98 \\ 
$n_1$-$n_2$+$g$+$\sigma$ & -1.2121926306         &                16461          & -70.50 \\ 
-$n_1$+$g$+$f$-$\sigma$ & -0.5151047705          &                15539          & 162.60 \\ 
3$n_2$-2$g$ & 1.0285105407       &                13249          & -167.30 \\ 
-$n_1$-5$n_2$+7$g$ & -0.8416386232       &                12481          & 162.43 \\ 
$f$+$\sigma$ & -1.3816866301     &                12264          & 173.83 \\ 
-2$n_2$+3$g$ & -0.6855437242     &                11888          & -63.52 \\ 
$f$-$\sigma$ & -1.3874424304     &                11817          & -1.08 \\ 
$n_1$+2$n_2$-$g$-$f$-$\sigma$ & 1.1951266386     &                11176          & -49.05 \\ 
-2$n_1$-3$n_2$+6$g$ & 0.7163207429       &                10310          & -75.41 \\ 
$n_1$+2$n_2$-$g$-$f$+$\sigma$ & 1.2008824403     &                10240          & 125.82 \\ 
-$n_1$+$g$+$f$-2$\sigma$ & -0.5179826716         &                 9898          & -14.83 \\ 
-10$n_1$-28$n_2$+39$g$ & -0.8752492838   &                 9724          & -46.99 \\ 
12$n_1$+28$n_2$-39$g$-2$\sigma$ & -0.8750258742          &                 9589   & 54.66 \\ 
-$n_1$+$g$+$f$+2$\sigma$ & -0.5064710674         &                 9447          & -25.11 \\ 
-$n_1$+$n_2$+$f$+$\sigma$ & -0.1665381171        &                 8998          & -43.25 \\ 
2$n_1$+$n_2$-$g$-$f$-2$\sigma$ & -0.0228997751   &                 8639          & 170.60 \\ 
-$n_1$+2$n_2$-2$\sigma$ & 1.5522815465   &                 8478          & -157.75 \\ 
-$n_1$+$n_2$+$f$-$\sigma$ & -0.1722939170        &                 8194          & 141.83 \\ 
2$n_1$+3$n_2$-4$g$ & -0.7161647792       &                 8079          & -134.65 \\ 
$n_1$-4$\sigma$ & -0.8837712815          &                 8017          & -78.45 \\ 
3$n_1$+5$n_2$-7$g$+$\sigma$ & -0.9000028324      &                 7864          & 17.59 \\ 
2$n_1$+$n_2$-$g$-$f$+2$\sigma$ & -0.0113881720   &                 7799          & 160.35 \\ 
$n_1$+2$n_2$-2$g$ & -0.1866379719        &                 7545          & -130.23 \\ 
3$n_1$+5$n_2$-7$g$ & -0.9028807332       &                 6818          & -159.85 \\ 
12$n_1$+28$n_2$-39$g$ & -0.8692700923    &                 6574          & 49.52 \\ 
-10$n_1$-28$n_2$+39$g$+2$\sigma$ & -0.8694934632         &                 6375   & -51.95 \\ 
$n_1$-2$n_2$+2$g$ & -1.5578813843        &                 6348          & 132.81 \\ 
$n_1$+2$n_2$-$g$-$f$-2$\sigma$ & 1.1922487381    &                 6203          & -46.49 \\ 
-2$n_1$+$n_2$+$g$+$f$-2$\sigma$ & 0.6971658421   &                 6038          & 128.07 \\ 
3$n_1$+4$n_2$-6$g$ & -1.2456915863       &                 5893          & -139.08 \\ 
\hline
\end{tabular}}\subfloat[$z=e \mathrm{exp}(i\varpi)$]{
\begin{tabular}{@{}C{3cm}R{2.5cm}R{1.3cm}r@{}}
\hline
 & $\nu_k$ (deg/day) & $10^6\times A_k$ & $\phi_k$ (deg) \\
\hline
$g$ & 0.0000779818       &               154688          & 74.97 \\ 
2$n_1$+5$n_2$-6$g$ & -0.0305430732       &                45211          & 3.83 \\ 
$n_2$ & 0.3428888349     &                42896          & 54.21 \\ 
-$n_1$+2$n_2$ & 1.5580373481     &                37713          & -162.87 \\ 
-$n_1$+2$n_2$+$\sigma$ & 1.5609152488    &                23004          & -165.43 \\ 
-$n_1$+2$n_2$-$\sigma$ & 1.5551594473    &                22252          & 19.69 \\ 
$n_1$ & -0.8722596782    &                18226          & 91.30 \\ 
2$n_2$-$g$ & 0.6856996878        &                17174          & -146.54 \\ 
$f$ & -1.3845645297      &                15986          & 176.36 \\ 
$n_1$-$\sigma$ & -0.8751375794   &                11183          & 93.87 \\ 
$n_1$+$\sigma$ & -0.8693817775   &                10857          & -91.26 \\ 
-$n_1$+$g$+$f$+$\sigma$ & -0.5093489691          &                10432          & -22.52 \\ 
$f$+$\sigma$ & -1.3816866296     &                 9721          & 173.82 \\ 
3$n_2$-2$g$ & 1.0285105408       &                 9583          & -167.30 \\ 
$n_1$+2$n_2$-2$g$ & -0.1866379723        &                 9567          & 49.78 \\ 
-$n_1$+$g$+$f$-$\sigma$ & -0.5151047704          &                 9500          & 162.60 \\ 
$f$-$\sigma$ & -1.3874424305     &                 9347          & -1.08 \\ 
-2$n_1$-3$n_2$+6$g$ & 0.7163207429       &                 8784          & -75.40 \\ 
$n_1$+$n_2$-$g$ & -0.5294488254          &                 8474          & 70.54 \\ 
2$n_1$+5$n_2$-6$g$-$\sigma$ & -0.0334209740      &                 7803          & -173.60 \\ 
-$n_1$+2$n_2$-2$\sigma$ & 1.5522815466   &                 6455          & -157.75 \\ 
-$n_1$+$n_2$+$f$+$\sigma$ & -0.1665381165        &                 6277          & -43.26 \\ 
-$n_1$+$g$+$f$-2$\sigma$ & -0.5179826711         &                 6048          & -14.84 \\ 
$n_1$+2$n_2$-2$g$-$\sigma$ & -0.1895158730       &                 5840          & 52.34 \\ 
$n_1$+2$n_2$-2$g$+$\sigma$ & -0.1837600719       &                 5837          & -132.77 \\ 
-$n_1$+$g$+$f$+2$\sigma$ & -0.5064710677         &                 5786          & -25.10 \\ 
-$n_1$+$n_2$+$f$-$\sigma$ & -0.1722939171        &                 5711          & 141.83 \\ 
$n_1$+$n_2$-$g$-$\sigma$ & -0.5323267261         &                 5228          & 73.11 \\ 
$n_1$+$n_2$-$g$+$\sigma$ & -0.5265709252         &                 5074          & -112.00 \\ 
-$n_1$+2$n_2$+2$\sigma$ & 1.5637931503   &                 4913          & -168.02 \\ 
$n_2$-$g$+$f$ & -1.0417536761    &                 4552          & 155.58 \\ 
4$n_2$-3$g$ & 1.3713213936       &                 4087          & 171.95 \\ 
-$n_1$+$n_2$+$f$-2$\sigma$ & -0.1751718176       &                 3618          & -35.61 \\ 
2$n_1$+5$n_2$-6$g$+$\sigma$ & -0.0276651723      &                 3602          & -178.74 \\ 
-2$n_1$-2$n_2$+5$g$ & 1.0591315959       &                 3549          & -96.16 \\ 
-2$n_1$+$n_2$+$g$+$f$+2$\sigma$ & 0.7086774454   &                 3522          & 117.82 \\ 
$n_1$+$n_2$-$f$-$\sigma$ & 0.8523157842          &                 3503          & -28.25 \\ 
-$n_1$+$n_2$+$f$+2$\sigma$ & -0.1636602134       &                 3481          & -45.89 \\ 
$n_1$+3$n_2$-3$g$ & 0.1561728811         &                 3433          & -150.98 \\ 
$n_1$-$n_2$+$g$ & -1.2150705304          &                 3399          & 112.03 \\ 
$n_1$+$n_2$-$f$+$\sigma$ & 0.8580715878          &                 3175          & 146.56 \\ 
-2$n_1$+$n_2$+$g$+$f$-2$\sigma$ & 0.6971658420   &                 3175          & 128.07 \\ 
$n_1$+2$\sigma$ & -0.8665038769          &                 3174          & 86.18 \\ 
-$n_1$+$g$+$f$ & -0.5122268693   &                 3071          & -19.98 \\ 
2$n_1$+$n_2$-$g$-$f$-2$\sigma$ & -0.0228997760   &                 2887          & -9.38 \\ 
-$n_1$+$n_2$+$g$ & 1.2152264949          &                 2877          & 37.89 \\ 
-$n_1$+$g$+$f$-3$\sigma$ & -0.5208605706         &                 2829          & 167.69 \\ 
$n_2$-$g$+$f$+$\sigma$ & -1.0388757781   &                 2774          & 153.11 \\ 
$f$-2$\sigma$ & -1.3903203311    &                 2688          & -178.52 \\ 
-2$n_1$+$n_2$+$g$+$f$+$\sigma$ & 0.7057995439    &                 2571          & 120.40 \\ 
\hline
\end{tabular}}
\caption{\label{tab:freq_planet}First 50 terms of the frequency decomposition $\sum_{k=1}^{50}A_ke^{i\left(\nu_k t+\phi_k\right)}$ of $\xi$ (a) and $z$ (b) for the planet on $\left[0:120\right]\mathrm{kyr}$ for the solution obtained for the best Newtonian fit for the coplanar case (Table~\ref{Table 2}).}
\end{table*}

\begin{table*}
\centering
\subfloat[$\xi=a \mathrm{exp}(i\lambda)$]{
\begin{tabular}{@{}C{3cm}R{2.5cm}R{1.3cm}r@{}}
\hline
 & $\nu_k$ (deg/day) & $10^6\times A_k$ & $\phi_k$ (deg) \\
\hline
$n_2$ & 0.3428888349     &              2610108          & 54.21 \\ 
$n_1$ & -0.8722596783    &                 5626          & 91.30 \\ 
$n_2$-$\sigma$ & 0.3400109341    &                 3923          & 56.78 \\ 
$n_2$+$\sigma$ & 0.3457667357    &                 3882          & -128.35 \\ 
$n_1$-$\sigma$ & -0.8751375791   &                 3480          & 93.86 \\ 
$n_1$+$\sigma$ & -0.8693817775   &                 3324          & -91.26 \\ 
-$n_1$+2$n_2$ & 1.5580373482     &                 1740          & -162.88 \\ 
-$n_1$+2$n_2$+$\sigma$ & 1.5609152488    &                 1064          & -165.44 \\ 
-$n_1$+2$n_2$-$\sigma$ & 1.5551594476    &                 1024          & 19.68 \\ 
$n_1$+2$\sigma$ & -0.8665038768          &                  963          & 86.18 \\ 
-$n_1$+$n_2$+$g$ & 1.2152264950          &                  759          & 37.89 \\ 
$n_1$-2$\sigma$ & -0.8780154800          &                  754          & 96.43 \\ 
$n_1$+3$n_2$-2$g$-$f$-$\sigma$ & 1.5379374918    &                  722          & -69.81 \\ 
$n_1$+$n_2$-$g$ & -0.5294488252          &                  680          & -109.46 \\ 
$n_1$+3$n_2$-2$g$-$f$+$\sigma$ & 1.5436932932    &                  656          & 105.07 \\ 
-$n_1$+$n_2$+$g$+$\sigma$ & 1.2181043959         &                  468          & 35.32 \\ 
-$n_1$+$n_2$+$g$-$\sigma$ & 1.2123485941         &                  446          & -139.55 \\ 
$n_1$-$n_2$+$g$ & -1.2150705314          &                  434          & 112.06 \\ 
$g$ & 0.0000779819       &                  424          & 74.97 \\ 
$n_1$+3$n_2$-2$g$-$f$+2$\sigma$ & 1.5465711935   &                  422          & -77.48 \\ 
$n_1$+$n_2$-$g$-$\sigma$ & -0.5323267261         &                  422          & -106.89 \\ 
$n_1$+3$n_2$-2$g$-$f$-2$\sigma$ & 1.5350595909   &                  404          & -67.25 \\ 
$n_1$+$n_2$-$g$+$\sigma$ & -0.5265709242         &                  403          & 67.97 \\ 
$n_1$+3$\sigma$ & -0.8636259768          &                  313          & -96.37 \\ 
-$n_1$+2$n_2$-2$\sigma$ & 1.5522815489   &                  296          & -157.82 \\ 
$n_1$-$n_2$+$g$-$\sigma$ & -1.2179484321         &                  270          & 114.62 \\ 
$n_1$-$n_2$+$g$+$\sigma$ & -1.2121926305         &                  255          & -70.51 \\ 
2$n_2$-$g$ & 0.6856996884        &                  236          & 33.44 \\ 
-$n_1$+2$n_2$+2$\sigma$ & 1.5637931500   &                  228          & -168.01 \\ 
-$n_1$-2$n_2$+5$g$-$f$+$\sigma$ & 1.5743143482   &                  217          & 176.21 \\ 
$n_1$+3$n_2$-2$g$-$f$ & 1.5408153924     &                  207          & -72.37 \\ 
$n_1$+3$n_2$-2$g$-$f$+3$\sigma$ & 1.5494490905   &                  200          & 100.06 \\ 
$n_1$-3$\sigma$ & -0.8808933805          &                  198          & 98.98 \\ 
$n_1$+2$n_2$-$g$-$f$-$\sigma$ & 1.1951266387     &                  181          & -49.05 \\ 
-$n_1$-2$n_2$+5$g$-$f$-$\sigma$ & 1.5685585468   &                  177          & 1.33 \\ 
-$n_1$+$g$+$f$+$\sigma$ & -0.5093489691          &                  176          & -22.52 \\ 
-2$n_1$-3$n_2$+6$g$ & 0.7163207426       &                  174          & -75.40 \\ 
$n_1$+2$n_2$-$g$-$f$+$\sigma$ & 1.2008824405     &                  165          & 125.82 \\ 
-$n_1$+$g$+$f$-$\sigma$ & -0.5151047706          &                  160          & 162.60 \\ 
-$n_1$-2$n_2$+5$g$-$f$+2$\sigma$ & 1.5771922489          &                  154   & -6.35 \\ 
$n_1$+3$n_2$-2$g$-$f$-3$\sigma$ & 1.5321816902   &                  150          & -64.69 \\ 
$n_2$+2$\sigma$ & 0.3486446368   &                  146          & 49.08 \\ 
$n_2$-2$\sigma$ & 0.3371330325   &                  143          & -120.64 \\ 
-$n_1$+$n_2$+$g$-2$\sigma$ & 1.2094706920        &                  129          & 43.05 \\ 
$n_2$-3$\sigma$ & 0.3342551310   &                  129          & 61.94 \\ 
$n_2$+3$\sigma$ & 0.3515225380   &                  125          & -133.49 \\ 
$n_1$+$n_2$-$g$+2$\sigma$ & -0.5236930221        &                  117          & -114.63 \\ 
-$n_1$-$n_2$+2$g$+$f$+$\sigma$ & -0.8521598223   &                  104          & -1.75 \\ 
-$n_1$+$g$+$f$-2$\sigma$ & -0.5179826722         &                  102          & -14.81 \\ 
$n_1$+2$n_2$-$g$-$f$-2$\sigma$ & 1.1922487378    &                  101          & -46.48 \\ 
\hline
\end{tabular}}\subfloat[$z=e \mathrm{exp}(i\varpi)$]{
\begin{tabular}{@{}C{3cm}R{2.5cm}R{1.3cm}r@{}}
\hline
 & $\nu_k$ (deg/day) & $10^6\times A_k$ & $\phi_k$ (deg) \\
\hline
$g$ & 0.0000779818       &               236253          & 74.97 \\ 
$n_1$ & -0.8722596783    &                 1783          & 91.30 \\ 
$n_2$ & 0.3428888349     &                 1113          & 54.21 \\ 
$n_1$-$\sigma$ & -0.8751375791   &                 1103          & 93.86 \\ 
$n_1$+$\sigma$ & -0.8693817775   &                 1054          & -91.26 \\ 
$n_1$+2$\sigma$ & -0.8665038766          &                  306          & 86.17 \\ 
$n_1$-2$\sigma$ & -0.8780154799          &                  239          & 96.43 \\ 
$n_1$+3$n_2$-2$g$-$f$-$\sigma$ & 1.5379374918    &                  219          & -69.81 \\ 
$n_1$+3$n_2$-2$g$-$f$+$\sigma$ & 1.5436932934    &                  199          & 105.06 \\ 
2$n_2$-$g$ & 0.6856996881        &                  146          & 33.45 \\ 
-$n_1$+2$n_2$ & 1.5580373479     &                  145          & 17.13 \\ 
-$n_1$+$n_2$+$g$ & 1.2152264951          &                  137          & -142.12 \\ 
$n_1$+3$n_2$-2$g$-$f$+2$\sigma$ & 1.5465711943   &                  128          & -77.50 \\ 
$n_1$+3$n_2$-2$g$-$f$-2$\sigma$ & 1.5350595909   &                  122          & -67.25 \\ 
2$n_1$+5$n_2$-6$g$ & -0.0305430732       &                  100          & 3.83 \\ 
$n_1$+3$\sigma$ & -0.8636259748          &                   99          & -96.42 \\ 
-$n_1$+2$n_2$+$\sigma$ & 1.5609152491    &                   89          & 14.56 \\ 
-$n_1$+$n_2$+$g$+$\sigma$ & 1.2181043958         &                   85          & -144.68 \\ 
-$n_1$+2$n_2$-$\sigma$ & 1.5551594463    &                   84          & -160.28 \\ 
-$n_1$+$n_2$+$g$-$\sigma$ & 1.2123485943         &                   81          & 40.45 \\ 
-$n_1$-2$n_2$+5$g$-$f$+$\sigma$ & 1.5743143484   &                   64          & 176.21 \\ 
$n_1$+3$n_2$-2$g$-$f$ & 1.5408153926     &                   63          & -72.37 \\ 
$n_1$-3$\sigma$ & -0.8808933806          &                   63          & 98.98 \\ 
$n_1$+3$n_2$-2$g$-$f$+3$\sigma$ & 1.5494490961   &                   60          & 99.91 \\ 
-$n_1$-2$n_2$+5$g$-$f$-$\sigma$ & 1.5685585467   &                   52          & 1.33 \\ 
-$n_1$-2$n_2$+5$g$-$f$+2$\sigma$ & 1.5771922492          &                   45   & -6.35 \\ 
$n_1$+3$n_2$-2$g$-$f$-3$\sigma$ & 1.5321816902   &                   45          & -64.68 \\ 
-$n_1$-$n_2$+2$g$+$f$+$\sigma$ & -0.8521598218   &                   44          & 178.23 \\ 
$n_1$+$n_2$-$g$ & -0.5294488251          &                   43          & -109.46 \\ 
-$n_1$-$n_2$+2$g$+$f$-$\sigma$ & -0.8579156257   &                   40          & 3.42 \\ 
$g$-$\sigma$ & -0.0027999189     &                   39          & 77.53 \\ 
-2$n_1$-3$n_2$+6$g$ & 0.7163207425       &                   35          & -75.39 \\ 
$n_1$+4$\sigma$ & -0.8607480674          &                   32          & 80.83 \\ 
2$n_1$+3$n_2$-3$g$-$f$-2$\sigma$ & 0.6627219308          &                   30   & 129.08 \\ 
-$n_1$-5$n_2$+7$g$ & -0.8416386234       &                   30          & -17.55 \\ 
$n_1$-2$n_2$+2$g$ & -1.5578813846        &                   29          & -47.18 \\ 
-$n_1$-2$n_2$+5$g$-$f$-2$\sigma$ & 1.5656806459          &                   29   & 3.90 \\ 
2$n_1$+3$n_2$-3$g$-$f$+2$\sigma$ & 0.6742335340          &                   28   & 118.83 \\ 
$n_1$+$n_2$-$g$-$\sigma$ & -0.5323267263         &                   26          & -106.89 \\ 
$n_1$+3$n_2$-2$g$-$f$+4$\sigma$ & 1.5523269947   &                   26          & -82.61 \\ 
-$n_1$+$n_2$+$f$+$\sigma$ & -0.1665381163        &                   25          & -43.27 \\ 
-$n_1$-$n_2$+2$g$+$f$-2$\sigma$ & -0.8607935214          &                   25   & -174.11 \\ 
$n_1$+$n_2$-$g$+$\sigma$ & -0.5265709240         &                   25          & 67.96 \\ 
-$n_1$-$n_2$+2$g$+$f$+2$\sigma$ & -0.8492819213          &                   24   & 175.68 \\ 
-$n_1$+2$n_2$-2$\sigma$ & 1.5522815372   &                   24          & 22.55 \\ 
-$n_1$+$n_2$+$g$-2$\sigma$ & 1.2094706942        &                   23          & -137.01 \\ 
-$n_1$+$g$+$f$+$\sigma$ & -0.5093489691          &                   23          & -22.52 \\ 
-$n_1$+$n_2$+$f$-$\sigma$ & -0.1722939171        &                   23          & 141.83 \\ 
-$n_1$-2$n_2$+5$g$-$f$+3$\sigma$ & 1.5800701501          &                   22   & 171.08 \\ 
2$n_1$+3$n_2$-3$g$-$f$+3$\sigma$ & 0.6771114347          &                   21   & -63.73 \\ 
\hline
\end{tabular}}

\end{table*}

\clearpage
\section{Additional figure}
\begin{figure*}
\includegraphics[width=\textwidth]{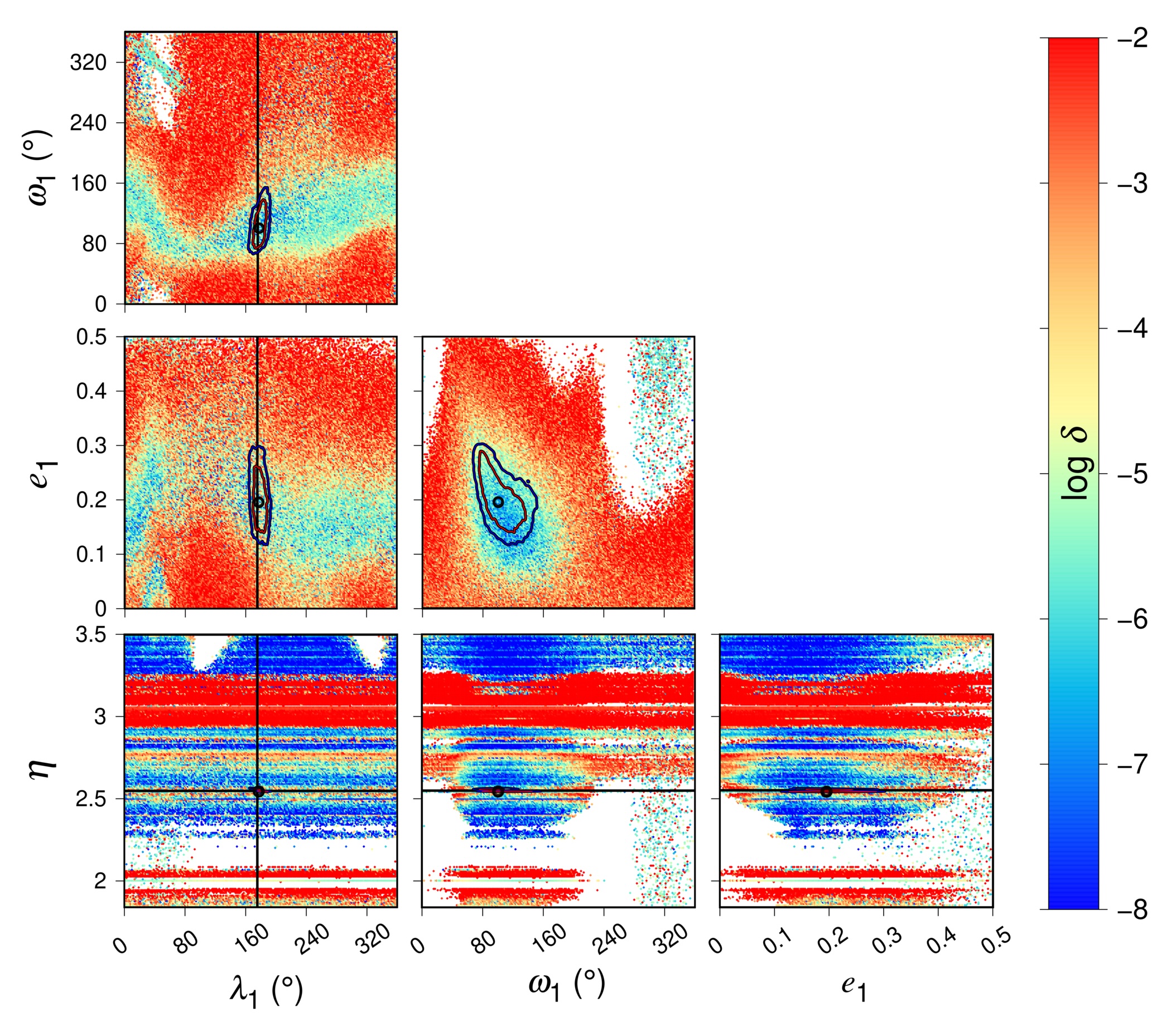}
  \caption{Corner Stability Maps considering all surviving initial conditions. The solid lines and the circles represent the Keplerian Fit and the Newtonian best stable fit, respectively. The confidence levels represent the $1\sigma$ (red) and $2\sigma$ (blue) regions. The colour scale corresponds to the diffusion index in logarithmic scale (Eq. \ref{diffusionindex}).}
  \label{CornerAll}
\end{figure*}

\end{document}